%% file: AdapMG.tex
\documentclass[preprint,12pt]{elsarticle}

\usepackage{amssymb}

\usepackage{mathrsfs}
\usepackage{graphicx}
\usepackage{amsmath}
\usepackage{amsfonts}
\usepackage{amsthm}
\usepackage{textcomp}
\usepackage{color}
\usepackage{xcolor}
\usepackage[normalem]{ulem}
\usepackage{placeins}
\usepackage{bm}
\usepackage{wasysym}
\usepackage{geometry}
\usepackage{stmaryrd}
\usepackage{float}
\usepackage{subfigure}
\usepackage{pdfsync}
\usepackage{algorithm, algorithmic}
\renewcommand{\algorithmiccomment}[1]{\bgroup\hfill//~#1\egroup}
\usepackage{dotlessi}

\newtheorem{theorem}{Theorem}[section]

\newtheorem{remark}{Remark}[section]

\renewcommand{\theequation}{\arabic{section}.\arabic{equation}}

\newcommand\R{\mathbb{R}}
\newcommand\Rg{\mathcal{R}}
\newcommand\T{\mathcal{T}}
\newcommand\No{\mathcal{N}}
\newcommand\N{\mathbb{N}}
\newcommand\Z{\mathbb{Z}}

\newcommand\E{\mathcal{E}}

\newcommand{\<}{\langle}
\renewcommand{\>}{\rangle}

\newcommand{\LO}{\L^{\Omega}}

\definecolor{yscol}{HTML}{6622AA}

\input{notation}

\begin{document}

\begin{frontmatter}

\title{Adaptive Multigrid Strategy for Geometry Optimization of Large-Scale Three Dimensional Molecular Mechanics}

\author[zjuaddress]{Kejie Fu\corref{equal}}
\cortext[equal]{~The first two authors contributed equally. ML's work was done primarily when he was a post-doctoral researcher at the Shanghai Jiao Tong University.}

\author[mingjieaddress]{Mingjie Liao\corref{equal}}

\author[yangshuaiaddress]{Yangshuai Wang\corref{mycorrespondingauthor}}
\cortext[mycorrespondingauthor]{~Corresponding author}
\ead{yswang2021@math.ubc.ca}

\author[zjuaddress]{Jianjun Chen}

\author[sjtuaddress]{Lei Zhang}

\address[zjuaddress]{Center for Engineering and Scientific Computation and School of Aeronautics and Astronautics, Zhejiang University, Hangzhou 310058, China.}

\address[mingjieaddress]{AsiaInfo Technologies Limited, Beijing 100193, China. \\
}

\address[yangshuaiaddress]{Department of Mathematics, University of British Columbia, Vancouver V6T1Z2, Canada.}

\address[sjtuaddress]{School of Mathematical Sciences, Institute of Natural Sciences and MOE-LSC, Shanghai Jiao Tong University, Shanghai 200240, China.}

\begin{abstract}
In this paper, we present an efficient adaptive multigrid strategy for the geometry optimization of large-scale three dimensional molecular mechanics. The resulting method can achieve significantly reduced complexity by exploiting the intrinsic low-rank property of the material configurations and by combining the state-of-the-art adaptive techniques with the hierarchical structure of multigrid algorithms. To be more precise, we develop a oneway multigrid method with adaptive atomistic/continuum (a/c) coupling, e.g., blended ghost force correction (BGFC)~\cite{colz2016} approximations with gradient-based {\it a posteriori} error estimators on the coarse levels. We utilize state-of-the-art 3D mesh generation techniques to effectively implement the method. For 3D crystalline defects, such as vacancies, micro-cracks and dislocations, compared with brute-force optimization, complexity with superior rates can be observed numerically, and the strategy has a five-fold acceleration in terms of CPU time for systems with $10^8$ atoms. 
\end{abstract}




\end{frontmatter}



\section{Introduction}
\label{sec:introduction}

In this paper, we develop a novel adaptive multigrid strategy for the geometry optimization of three dimensional molecular mechanics, by utilizing the intrinsic low-rank property of materials configuration and a judicious combination of the adaptive and hierarchical algorithmic structure, and in addition, a crystalline structure adapted three dimensional mesh generation technique.

Geometry optimization is referred to as the process to locate the minimum energy configuration of materials and molecules, and it is one of the most crucial and rate determining steps of computational materials and molecular science \cite{van2020roadmap, tadmor2011modeling,  tadmor2000hierarchical}. The design and analysis of efficient geometry optimization algorithms have attracted considerable attentions from both the engineering community and the mathematical community in recent years \cite{bisbo2020efficient, bitzek2006structural, chen2017efficient, chen2011efficient, mones2018preconditioners, wales2003energy}. 
Standard optimization techniques such as line-search-based descent schemes like steepest descent or conjugate gradient \cite{hager2006algorithm, ruder2016overview}, pseudo-dynamics relaxation schemes like the fast inertial relaxation engine (FIRE, implemented in LAMMPS \cite{bitzek2006structural, guenole2020assessment}) and Quasi-Newton schemes like L-BFGS \cite{liu1989limited} are widely used in atomistic simulations. We call the above mentioned methods {\it brute-force} optimization as their computational complexity is at least linearly proportional to the number of atoms ({\it dofs}). More seriously, complex defects such as dislocations and cracks induce a long-range elastic field, which always leads to a super-linear complexity in terms of the {\it dofs} when {\it brute-force} optimization is employed, see e.g., Figure \ref{crack-cpu-time} for the micro-crack case and Figure \ref{dislocation-cpu-time} for the dislocation case.
To achieve better performance for geometry optimization tasks, we need to explore the intrinsic structure of the materials configuration from the modeling perspective, and to combine the state-of-the-art optimization techniques from the algorithmic perspective.

From the modeling perspective, it is well-known that materials can be described by a hierarchy of coarse grained models with due accuracy. For materials with no defects and relatively smooth deformation, Cauchy-Born model provides a second order approximation and allows for a coarse discretization \cite{MB_KH_Crystal_Latices, e2007cb, ortner13}. For materials with localized defects, atomistic/continuum (a/c) modeling takes a domain-decomposition formulation such that the atomistic models are applied in a small neighborhood of the localized defects while continuum models are employed away from the defect cores. We refer to \cite{van2020roadmap, luskin2013atomistic, miller2009unified, olson2014optimization, olson2016analysis} for reviews of many existing a/c coupling methods. A/c coupling methods have sublinear computational cost due to its adaptive nature, but its efficient implementation relies on the development of efficient and robust a posteriori estimators and sophisticated mesh generations techniques which can be adapted to the underlying crystalline structure \cite{phlipot2019quasicontinuum, 2006_SP_PB_TO_IJMCE, tembhekar2017automatic, xu2019modeling}, especially in three dimensions. In recent works \cite{2021-defectexpansion, 2013-defects}, low rank structures of defect configurations can be revealed by a novel far-field expansion of the long-range elastic fields, which in turn provides a good {\it predictor} of the atomistic equilibrium. 

From the algorithmic perspective, the atomistic model and the coarse grained models such as Cauchy-Born and a/c coupling are highly nonconvex, which require sophisticated optimization methods. The multigrid method, originally introduced as a scalable linear solver  \cite{brandt1977multi, hackbusch2013multi}, can be applied on the non-linear optimization with linear complexity \cite{gratton2008recursive, nash2000multigrid, wen2010line}. The application of multigrid methods in molecular mechanics is originated by \cite{brandt1977multi}. Chen et. al. \cite{chen2011efficient, chen2014constrained} proposed a more general approach, where the coarse-grid operator is constructed by using the Cauchy–Born rule \cite{e2007cb, ortner13}, though its efficiency might be hindered by the reduced accuracy in the defect core region. The so-called quasi-atomistic approximation is proposed in \cite{chen2017efficient}, while its computational complexity still scales linearly with respect to the atomistic {\it dofs} on each coarse-grid level. 

In this paper, we aim to combine the adaptive nature of material modeling and hierarchical structure of multigrid algorithm to develop an efficient multigrid strategy (concluded as Algorithms \ref{alg:oneway}, \ref{alg:refine} and \ref{alg:bgfc}) and to achieve the optimal computational cost for the geometry optimization of molecular mechanics. We note that adaptive multigrid methods have been constructed in~\cite{bai1987local, brezina2006adaptive, rude1993fully} for the efficient solution of partial differential equations. We exploit the fact that the coarse grained models with sublinear complexity reveal the low-rank structure of the material configurations and can therefore be utilized as the coarse-grid problems in an adaptive and hierarchical manner. Therefore, a sequence of coarse-grid problems can be solved adaptively and can provide a sufficiently good initial guess for the geometry optimization on the finest level. Our main contributions can be summarized in the following.

Firstly, our method can achieve atomistic accuracy with significantly reduced complexity as the coarse operator in the multigrid scheme is based on the a/c coupling and involves only coarse {\it dofs}. In particular, we use the blended ghost force correction (BGFC) method \cite{colz2016} which integrates two popular ideas: blending \cite{luskin2013formulation} and ghost force correction \cite{Shenoy:1999a}. It is easy to implement and achieves the optimal convergence rate among all a/c coupling employing Cauchy-Born model in the continuum region. 

Secondly, we introduce adaptivity in each coarse level of the multigrid scheme, which results in an automatic partition of the atomistic and continuous subsystems as well as an adaptive local mesh refinement in the continuum region, and in turn achieves a (quasi-)optimal balance between accuracy and efficiency at each level of the multigrid iteration. The design of robust and efficient a posteriori error estimator is the key to accomplish this. Heuristic methods have been proposed in the engineering literature \cite{tembhekar2017automatic, Shenoy:1999a}, while the development of {\it a posteriori} analysis and adaptive algorithm for a/c coupling methods in two or three dimensions are quite recent. We refer to our works \cite{liao2018posteriori, wang2018posteriori} in this direction, where the residual-based error estimators based on the stress tensor formulation are analyzed. In this work, we use the gradient-based error estimator for the sake of simplicity. 

The last but not least, our work provides a 3D implementation for complex defect configuration, using crystalline structure adapted mesh generation. The 3D mesh generation and adaptation are crucial for the effective implementation of a/c coupling simulations. Compared to the relevant works in two dimensions \cite{tembhekar2017automatic, wang2018posteriori}, the extension to three dimensions is highly nontrivial: (i) the atomistic region is not guaranteed to be convex and the surface mesh needs to be constructed; (ii) a smooth transition region is required for complex defects with large distortions; (iii) a robust mesh adaptation is essential in three dimensions. We leave the detailed constructions and discussions in a separate work \cite{CPCmesh} and mainly focus on its applications to the subject of this paper.

We implement the main algorithm and test several prototypical benchmark examples of crystalline defects such as single-vacancy, micro-crack and edge dislocation in three dimensions. The resulting adaptive multigrid strategy proposed in this paper can achieve significantly reduced complexity, which is much faster than the {\it brute-force} optimization and the quasi-atomistic approximation in \cite{chen2017efficient} with adaptive local mesh refinement. For systems with up to a hundred million atoms, our proposed strategy has a five-fold acceleration in terms of CPU time compared with the {\it brute-force} optimization. We plan to explore the generalization of this work in the future, including the more complex multigrid strategies (e.g., the full multigrid (FMG) and the full approximation scheme (FAS)) and the extension to the realistic crystalline defects such as partial dislocations and grain boundaries. 


\subsubsection*{Outline}
The paper is organized as follows: In \S~\ref{sec:ms}, we first introduce the atomistic model and the geometry optimization problem, and then formulate quasi-atomistic \cite{chen2017efficient} and BGFC \cite{colz2016} coarse models. In \S~\ref{sec:numerics}, we design the adaptive multigrid algorithms for the large-scale molecular mechanics geometry optimization using these coarse-grid models. We implement and test the algorithms for several typical crystalline defects in \S~\ref{sec:ne}. Future developments of the method are discussed in \S~\ref{sec:conclusion}.

\subsubsection*{Notation}
We use the symbol $\<,\>$ to denote an abstract duality pair between a Banach space $V$ and its dual space $V^{*}$.
The closed ball with radius $r$ and center $x$ is denoted by $B_r(x)$, and
$B_r:=B_r(0)$. For a finite set $A$, we will use $\#A$ to denote the cardinality of $A$.
%

\section{Model Setup}
\label{sec:ms}

In this section, we first set up the atomistic problem for crystalline defects. For the sake of simplicity, we consider the single-species Bravais lattices, and we note that all the algorithms discussed in this work can be applied to {\it multilattice} crystals \cite{olson2019theoretical}. 
To accelerate geometry optimization of the atomistic problem, we introduce two kinds of coarse-grid problems, namely, the quasi-atomistic (QA) approximation \cite{chen2017efficient} and the blended ghost force correction (BGFC) approximation~\cite{colz2016} using atomistic-to-continuum coupling.


\subsection{Atomistic problem}
\label{sec:ap}

\def\Rcore{R_{\rm DEF}}
\def\UsH{{\mathscr{U}}^{1,2}}
\def\Adm{{\rm Adm}}

Let $\Lhom=\mA\Z^3$ with some non-singular matrix $\mA\in\R^{3\times 3}$ be a perfect single lattice possessing no defects and $\L\subset \R^3$ be the corresponding single lattice with some local defects. The mismatch between $\L$ and $\Lhom$ 
represents possible defects containing some localized defect cores. 
Without loss of generality, for the case of a single defect, we assume that it is contained within a ball $B_{\Rcore}$ for $\Rcore>0$; that is, $\L \cap B_{\Rcore}$ is finite and $\L \setminus B_{\Rcore} = \Lhom \setminus B_{\Rcore}$. The case with multiple local defects can be similarly defined.  

We denote by $\Us := \{v: \L\to \mathbb{R}^3 \}$ the set of vector-valued lattice functions. Recall that the deformed configuration of $\L$ is a map $y\in \Us$ which can be decomposed as
\begin{eqnarray}\label{y-u}
y(\ell) = \ell + u_0(\ell) + u(\ell) \qquad\forall~\ell\in\Lambda,
\end{eqnarray}
where $u_0: \L\rightarrow\R^3$ is a {\it far-field predictor} enforcing the presence of the defect of interest and $u: \L\rightarrow\R^3$ is a {\it corrector}.
For point defects, we simply take $u_0=0$. For straight dislocations, $u_0$ can be derived by solving a continuum linearized elasticity (CLE) equation and we refer to~\cite{2021-defectexpansion, 2013-defects} for more details.  
%

For each atom $\ell\in \L$, we define the finite difference stencil for $v \in \Us$
\begin{align*}
Dv(\ell):= \{D_\rho v(\ell)\}_{\rho \in \Rg_\ell} :=\{v(\ell+\rho)-v(\ell)\}_{\rho \in \Rg_\ell},
\end{align*} 
where $\Rg_\ell := \{\ell'-\ell~|~\ell'\in \Nhd_\ell\}$ is the interaction range with interaction neighborhood $\Nhd_{\ell} := \{ \ell' \in \L~|~0<|\ell'-\ell| \leq \rcut \}$ with some cut-off radius $\rcut>0$.

\def\Use{\Us^{1,2}}
To measure the local ``regularity" of a displacement function $u \in \Us$, it is convenient to use a background mesh, for example, the {\it canonical} tetrahedral mesh $\T_{\L}$ of $\R^3$, whose nodes are the reference lattice points in $\L$.
We define $I u$ as the standard piecewise affine interpolation of $u$ with respect to $\T_{\L}$. When no confusion arises, we identify $u=Iu$ and then denote the piecewise constant gradient $\nabla u := \nabla I u$. We then introduce the discrete homogeneous Sobolev spaces via the nodal interpolant \cite{colz2016, 2013-defects, 2014-bqce}
\begin{displaymath}
	\Use :=\{u\in \Us ~|~\nabla u\in L^2\},
\end{displaymath}
with semi-norm $|\cdot|_{\Use}:=\|\nabla u\|_{L^2}$.

The site potential is a collection of mappings $V_{\ell}:(\R^3)^{\Rg_\ell}\rightarrow\R$, which represents the energy distributed to each atomic site in $\L$. We refer to \cite[\S~2]{2013-defects} and \cite[\S~2]{chen19} for the detailed discussions on the assumptions of general site potentials. In this work, we will use the well-known EAM (Embedded Atom Method) model \cite{Daw1984a} throughout the numerical experiments (cf. \S~\ref{sec:ne}).

We can formally define the energy-difference functional of the atomistic model
\begin{align}
\label{energy-difference}
\E(u) =& \sum_{\ell\in\Lambda}\Big(V_{\ell}\big(Du_0(\ell)+Du(\ell)\big)-V_{\ell}\big(Du_0(\ell)\big)\Big).
\quad
\end{align}

An equilibrium defect geometry is obtained by solving the following minimization problem
\begin{eqnarray}\label{eq:variational-A-problem}
u^{\rm a} \in \arg \min \big\{\E(u) ~\big|~ u \in \Use \big\}.
\end{eqnarray}
where ``$\arg\min$'' is understood as the set of local minima. It was shown in~\cite{2013-defects} that $\E(u)$ is well-defined, namely, the solutions to \eqref{eq:variational-A-problem} exist under suitable assumptions (cf.~\cite[\S~2.1]{2013-defects}). 

For meaningful numerical approximations, we need to project the atomistic problem to a finite-dimensional subspace. Due to the decay estimates of the energy minimizer~\cite[Theorem 1]{2013-defects}, it is natural to restrict the infinite lattice $\L$ to a finite domain. We choose a finite computational domain $\Omega_R \subset \R^3$ satisfying $B_{R^{\rm i}} \subset \Omega_R \subset B_{R^{\rm o}}$ with suitable constants $0<R^{\rm i}<R<R^{\rm o}$. When there is no ambiguity, we use $\Omega$ instead of $\Omega_R$ for simplicity. Let $\L^{\Omega}:=\L \cap \Omega$ and $N:=\#\L^{\Omega}$, we can modify the displacement space $\Use$ as
\begin{displaymath}
	\Use_N :=\big\{u\in \Use ~\big|~ u=0 ~\text{in}~ \L \setminus \L^{\Omega}\big\},
\end{displaymath}
where the clamped boundary condition is applied. We now formulate the approximate atomistic problem as
\begin{equation}\label{eq::a}
\mathbf{P}^{\rm A}_{N}: \quad u^{\rm a}_N \in  \arg\min \big\{ \E(u) ~\big|~ u \in \Use_N \big\},
\end{equation}
where ``$\arg\min$'' is understood as the set of local minima. In this work, we focus on the large-scale (e.g. $N>10^5$) molecular geometry optimization in three dimensions.

The computational complexity to solve \eqref{eq::a} is at least linear with respect to $N$ when {\it brute-force} optimization is utilized. To accelerate the geometry optimization of large-scale atomistic simulations, motivated by the sparse/adaptive representations of the defect configurations, we will introduce two coarse-grid problems in the following sections. 

\subsection{Quasi-atomistic (QA) approximation}
\label{sec:qa}

Given $k\in \mathbb{N}$, let $\T_{k}$ be the $k$-th level shape-regular simplicial partition of $\Omega$ (for example, obtained by refining $k$ times from some initial partition $\T_0$) and $\No_{k}:=\{\xi_1, ..., \xi_{n_k}\}$ be the set of all nodes of $\T_{k}$, where $n_k:=\#\No_k$. We note that the coarse-grid points ${\xi_1, ..., \xi_{n_k}}$ may not necessarily be the atomistic reference positions $x(\ell)$, for $\ell \in \LO$. 

We define $\psi\big(\cdot-\xi_j\big)$ as the basis function with compact support, and centered on the $j$th node of the given partition $\T_k$. The basis function $\psi$ is usually chosen such that the partition of unity condition 
\begin{equation}\label{eq::weq}
\sum^{n_k}_{j=1} \psi\big(\cdot-\xi_j\big) = 1, 
\end{equation}
is satisfied. A possible choice of $\psi$ is the standard piecewise affine basis function. 

On the $k$-th level, given $u_k:\No_k \to \R^3$, for any $\ell\in\L^{\Omega}$, the atomistic displacement at site $\ell$ can be approximated by
\begin{equation}\label{eq::interpu}
u(\ell) = \sum_{j=1}^{n_k} \psi\big(\ell-\xi_j\big) u_k(\xi_j).
\end{equation}
By inserting \eqref{eq::interpu} into \eqref{energy-difference}, we can obtain the energy functional of the quasi-atomistic (QA) approximation~\cite{chen2017efficient}
\begin{equation}\label{eqn:Eqa}
  \E^{\rm QA}(u_k) := \E \left(\sum_{j=1}^{n_k} \psi\big(\ell -\xi_j\big)u_k(\xi_j)\right).
\end{equation}

The coarse-grid problem for the QA approximation on the $k$-th level reads,
\begin{equation}\label{eq::uka}
\mathbf{P}^{\rm QA}_k: \quad
u^{\rm qa}_k \in \arg\min \{ \E^{\rm QA}(u_k) ~\big|~ u_k \in \Us^{\rm QA}_k \},
\end{equation}
where the solution space on the $k$-th level is defined by 
\begin{align}
  \Us^{\rm QA}_{k} := \big\{ u_k : \No_k \to \R^d ~\big|~
  \text{ $\nabla u \in L^2$, }
  \text{ $u_k = 0$ on $\partial \Omega$ } \big\}.
\end{align}

In \eqref{eqn:Eqa}, the contribution of each atom is explicitly accounted while only the displacements at the grid points are treated as unknowns in the coarse-grid optimization problem. Therefore, the computational complexity of solving \eqref{eq::uka} on the $k$-th level is $\mathcal{O}(N+n_k)$. 

\begin{remark}
From the above discussion, the computational complexity for the quasi-atomistic approximation on the coarse level is prohibitive for large scale simulations as it depends on the atomistic \textit{dof} $N$ linearly, where $N:=\#\L^{\Omega}$.
If we want to achieve sublinear complexity with respect to $N$ on each coarse level, we can exploit the low-rank structure of material configurations by using, e.g. Cauchy-Born model~\cite{e2007cb} or atomistic-to-continuum (a/c)~\cite{luskin2013atomistic} model. We will introduce the BGFC model, a typical a/c coupling model as the coarse problem in the next section. 
\end{remark}


\subsection{BGFC Method}
\label{sec:bgfc}

In this section, we first briefly review the well-known Cauchy-Born continuum model, then introduce the blended ghost force correction (BGFC) method which couples the atomistic model and Cauchy-Born continuum model. We keep the presentation as concise as possible and refer to \cite{colz2016, 2014-bqce, fang20} for more details. 

\subsubsection{Continuum approximation}
\label{sec:ca}
A continuum model can be derived by coarse-graining from the atomistic model \eqref{energy-difference}. Generally speaking, it allows for the reduction of {\it degrees of freedom} ({\it dofs}) and still keep sufficient accuracy when the deformation is smooth, e.g. the region far away from the defect core. Cauchy-Born continuum model~\cite{e2007cb, ortner13} is a typical choice in the multi-scale context. Let $W : \R^{3 \times 3} \to \R$ be a strain energy density function, the Cauchy-Born energy density $W$ is defined by 
\begin{displaymath}
  W(\mF) := V( \mF \cdot \Rg),
\end{displaymath}
and the Cauchy-Born energy difference reads
\begin{eqnarray}
W'(\mF) := W(\mF+{\bf 0}) - W({\bf 0}), \quad \forall \mF \in \R^{3\times 3},
\end{eqnarray}
where $\mI \in \R^{3\times 3}$ is the identity matrix. 

We note that the Cauchy-Born approximation can not capture the microscopic behaviour in the defect core,
therefore it is not a good choice for a coarse grained model of material defects. It slows down the convergence process in the multigrid strategies in \cite{chen2011efficient, chen2014constrained}. Instead, in the next section, we introduce the blended ghost force correction (BGFC) method, which will be used as the coarse-grid problem to achieve sublinear computational complexity with (quasi-)optimal accuracy~\cite{colz2016}.

\subsubsection{BGFC method}
\label{sec:bgfcm}

The a/c coupling methods is a class of concurrent multiscale methods which hybridizes atomistic and continuum models and can achieve an optimal accuracy with sublinear complexity. We refer to~\cite{van2020roadmap, miller2009unified} for the extensive overview and benchmark of a/c coupling methods for material defect simulation and~\cite{luskin2013atomistic} for their rigorous analysis.
In this work, we adopt the BGFC method as the coarse problem in our multigrid strategy, and the motivation is twofold: (i) it combines the benefits of blending~\cite{2014-bqce} and ghost force correction~\cite{Shenoy:1999a}, which is easy to implement, especially for three dimensional problems; (ii) it has the optimal rates of convergence, in terms of the number of degrees of freedom, among all a/c coupling methods employing Cauchy-Born model in the continuum region \cite{colz2016}. 

\begin{figure}[htb]
	\centering 
	\includegraphics[height=7cm]{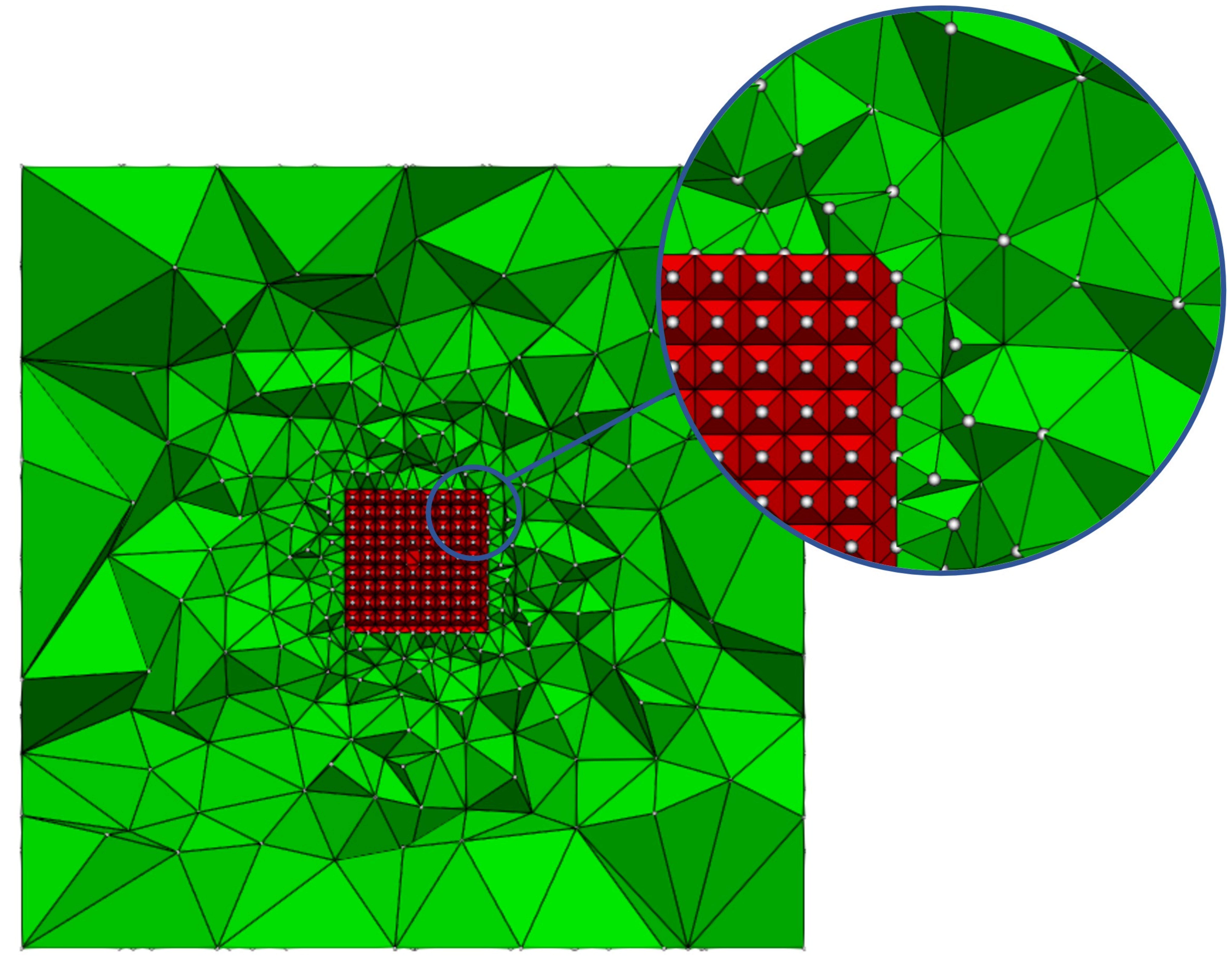}
	\caption{Illustration of the decomposition of $\T_k$ (2D slice), where $\T^\a_{k}$ is colored red while $\T^\c_{k}$ is colored green. The white dots in the red region represent atoms belonging to $\L^{\a}\cup\L^{\b}$ and others are the nodes of $\T^\c_k$. The atomistic region contains those atoms near the point defect. The graded tetrahedral mesh in the continuum region is constructed from the atomistic scale to the boundary length scale.
}
	\label{fig:acmesh}
\end{figure}

To construct the BGFC method, we first decompose the computational domain $\Omega = \Omega^\a \cup \Omega^\b \cup \Omega^\c \subset \R^3$ into three regions, the {{\it atomistic region}} $\Omega^{\a}$ with radius $R^\a$, the {{\it blending region}} $\Omega^\b$ with width $L^{\b}$ \footnote{To achieve the optimal rates of convergence, according to the {\it a priori} analysis given in \cite{2014-bqce}, the width of the blending region $\Omega^{\rm b}$ is fixed as $L^{\rm b} = R^\a$ throughout this paper. 
} 
and the {{\it continuum region}} $\Omega^\c$. Given the reference lattice $\Lambda$ with some local defects, we define the set of core atoms $\L^{\a} := \L \cap \Omega^\a$ and the set of blending atoms $\L^\b := \L \cap \Omega^\b$. Let $\T^\a_{k}$ be the {\it canonical} tetrahedral mesh induced by $\L^{\a}\cup\L^\b$, and $\T^{\rm c}_{k}$ be a shape-regular tetrahedral partition of the continuum region. We denote $\T_k = \T^\a_{k} \bigcup \T^\c_{k}$ as the tetrahedral partition on the $k$-th level. See Figure \ref{fig:acmesh} for an illustration of $\T_k$, where the construction details will be given in~\cite{CPCmesh}. 

We introduce the blending function $\beta \in C^{2,1}(\R^d)$ with $\beta=0$ in $\Omega^\a$ and $\beta=1$ in $\Omega^\c$. 
%
For the simplicity of presentation and implementation, we use $\mathcal{P}_1$ finite element to discretize the Cauchy-Born continuum model in the BGFC method, though $\mathcal{P}_2$ elements can be applied to achieve higher order convergence rate \cite{colz2016, fang20}. The space of coarse-grained displacements is therefore given by
\begin{align*}
  \Us^{\rm BGFC}_{k} := \big\{ u_k : \Omega \to \R^3 ~\big| ~&
  \text{ $u_k$ is continuous and p.w. affine w.r.t. $\T_{k}$, } \\[-1mm]
  & \text{ $u_k = 0$ in $\L \setminus \Omega$ } \big\}.
\end{align*}

The BGFC energy functional is based on a {\it second renormalization} of the potential \cite{colz2016}, for $\ell \in \L$, $u \in \Use$,
\begin{eqnarray}
V''_{\ell}(Du):=V_{\ell}(Du+Du_0)-V_{\ell}(Du_0)-\<\delta V_{\ell}(Du_0), Du\>,
\end{eqnarray}
where $u_0$ is the {\it predictor} introduced in \S~\ref{sec:ap}.
The corresponding second renormalized Cauchy-Born energy density is
\begin{eqnarray}
W''(\mF) := W(\mF+{\bf 0}) - W({\bf 0}) - \partial W({\bf 0}):\mF, \quad \text{for}~\mF\in \R^{3\times 3}.
\end{eqnarray}

Let $Q_{\rm h}$ be the $\mathcal{P}_0$ midpoint interpolation operator, such that $\int_{\Omega}Q_{\rm h} f \dx$ is the mid-point quadrature rule of $\int_{\Omega}f \dx$. We obtain the following BGFC energy functional \cite{colz2016},
\begin{align}\label{eqn:Eqc}
  \E^{\rm BGFC}(u_k) =& \sum_{\ell \in \L^{\a} \cup \L^{\b}} (1-\beta(\ell))V''_\ell(Du_k) + \int_{\Omega}Q_{\rm h}\beta(x)\Big[ W''(\nabla u_k + \nabla u_0)-W''(\nabla u_0) \Big]\dx \nonumber \\
  &\quad + \<L^{\rm ren}, u_k\>,
\end{align}
where the renormalization operator $L^{\rm ren}$ is defined by
\begin{align}\label{eq:ren}
\<L^{\rm ren}, u_k\> := \sum_{\ell\in\L\cap B_{\Rcore}} \big\<\delta V_{\ell}(Du_0), Du_k \big\> - \sum_{\ell \in \Lhom\cap B_{\Rcore}} \big\< \delta V({\bf 0}), D\tilde{u}_k\big\> 
\end{align}
with $\tilde{u}_k$ an arbitrary extension of $u_k$ from $\L$ to $\Lhom$ \cite[Appendix A]{wang2018posteriori}.

The coarse-grid problem for the BGFC method on the $k$-th level reads
\begin{equation}\label{eq::pbgfc}
\mathbf{P}^{\rm BGFC}_k: \qquad 
u^{\rm bgfc}_k \in \argmin \big\{ \E^{\rm BGFC}(u_k) ~\big|~ u_k \in \Us^{\rm BGFC}_k \big\}. 
\end{equation}
the computational complexity of solving \eqref{eq::pbgfc} is then proportional to  $n_k$, the {\it dofs} on the $k$-th level. It is much cheaper to solve compared with the reference geometry optimization problem \eqref{eq::a} since $n_k \ll N$. Hence, it is possible to obtain an efficient (sublinear complexity with respect to $N$) multigrid strategy when the BGFC method is employed as the coarse-grid problem.

The following theorem gives the {\it a priori} error estimates of BGFC method in terms of the computational cost $n_k$ for point defects in three dimensions and anti-plane screw dislocation, which follows from \cite[\S~4.2.1]{colz2016} and \cite[Theorem 2.1]{fang20} respectively.  

\begin{theorem}\label{thm:bgfc}
Suppose that the blending function $\beta$ and the triangulation $\T_k$ satisfy \cite[Assumption 1]{fang20}, and $\mathcal{P}_1$ finite element method is applied in the continuum region, for point defects in three dimensions, we have
\[
\|\nabla u^{\rm a} - \nabla u^{\rm bgfc}_k\|_{L^2} \lesssim n_k^{-5/6}.
\] 
For anti-plane screw dislocation with nearest-neighbour interaction, we can obtain
\[
\|\nabla u^{\rm a} - \nabla u^{\rm bgfc}_k\|_{L^2} \lesssim n_k^{-1} (\log n_k)^{1/2},
\]
where $u^{\rm a}$ and $u^{\rm bgfc}_k$ are the solutions of \eqref{eq:variational-A-problem} and \eqref{eq::pbgfc} respectively.
\end{theorem}

\begin{remark}\label{re:disloc}
We admit that the {\it a priori} error estimate of BGFC method for general straight dislocations is still lacking due to the evaluation of \eqref{eq:ren} is non-trivial (see \cite[Section 4.2]{colz2016} for a detailed discussion). A possible approach is to study the equivalent ghost force removal formulation \cite{Shenoy:1999a}
\begin{eqnarray}\label{eq:gfc}
\E^{\rm bgfc}(u_k) = \E^{\rm bqce}(u_k) - \big\<\delta \E^{\rm bqce}_{\rm hom}(\hat{u}_0), Du_{k}\big\>,
\end{eqnarray}
where $\E^{\rm bqce}_{\rm hom}(\hat{u}_0)$ is the energy-based blended quasi-continuum (BQCE) energy functional at a suitable ``predictor" $\hat{u}_0$. In our implementation, we apply \eqref{eq:gfc} instead of \eqref{eqn:Eqc} to formulate the BGFC method, where we simply choose $\hat{u}_0=0$ for all types of crystalline defects considered in this paper. We will explore this alternative point of view rigorously in future work, in particular with an eye to nontrivial choices of $\hat{u}_0$ in applications involving cracks and dislocations.
\end{remark}

\subsubsection{Adaptive BGFC method}

A fundamental challenge for a/c coupling methods is to optimally assign the atomistic/continuum regions and determine the mesh structure so that a (quasi-)optimal balance between accuracy and efficiency can be achieved. {\it A priori} choices (e.g. Theorem~\ref{thm:bgfc}), even though they are feasible, typically lead to sub-optimal distribution of computational resources and only work for simple setups such as single point defect. Hence, the {\it a posteriori} analysis and corresponding adaptive algorithms play a crucial role in the efficient implementation of a/c coupling methods.

The key to the {\it a posteriori} analysis for a/c coupling methods is to prove the following estimate
\[
\|\nabla u^{\rm a} - \nabla u^{\rm bgfc}_k\|_{L^2} \lesssim \eta(u^{\rm bgfc}_k),
\]
where $\eta(u^{\rm bgfc}_k)$ is called the {\it a posteriori} error estimator. It can be further assigned to local contributions, which gives instructions on how to move the a/c interface and adjust the
discretization of the continuum region automatically.
We refer to our recent works \cite{liao2018posteriori, wang2018posteriori, CMAME, wang2021posteriori} for more details in this direction, for example, the residual-based error analysis based on the stress tensor formulation, and the extensions on adaptive QM/MM coupling methods. 

Since our main interest in this paper is to develop a multigrid strategy by utilizing the idea of adaptivity, for simplicity, we use the heuristic gradient-based error estimator introduced in \S~\ref{sec:algo} throughout this paper. A rigorous {\it a posteriori} error estimate for the BGFC method will be investigated in our future work.



\section{Numerical Algorithms}
\label{sec:numerics}

We present the main adaptive multigrid strategy in this section. We first introduce the mesh generation which adapts to the crystalline structure and a/c coupling schemes in \S~\ref{sec:at}, and the oneway multigrid strategy in \S~\ref{sec:mstra}, respectively. We then design the main algorithm which employs the BGFC coarse-grid problem in \S~\ref{sec:algo}. For simplicity, we leave the algorithm with quasi-atomistic (QA)  coarse-grid problem in the~\ref{sec:appendix}.

\subsection{Mesh generation}
\label{sec:at}

Mesh generation and mesh adaptation play an important role in the multiscale modeling and simulations. The performance of adaptive algorithms are heavily influenced by its quality and efficiency.
In this paper, on each coarse-grid level, we use two different partitions for two coarse-grid problems. More precisely, the tetrahedral mesh is used in the continuum region for the BGFC method while the cubic mesh is applied for the QA approximation which is introduced in the~\ref{sec:appendix}. 

In the context of a/c coupling methods \cite{ortner13, miller2009unified}, the tetrahedral mesh is more natural than the cubic mesh since it is more natural to be consistent with the underlying crystalline structure, especially at the a/c interface, and the strain ($\nabla u_k$) evaluation in the Cauchy-Born energy density \eqref{eqn:Eqc} is more straightforward, which is also shown in~\cite{tembhekar2017automatic, radovitzky2000tetrahedral} from an engineered standpoint.
Hence, we obtain the tetrahedral mesh $\T^{\rm c}_k$ in the continuum region when the BGFC problem $\mathbf{P}^{\rm BGFC}_k$ defined by \eqref{eq::pbgfc} is considered as the coarse-grid problem on the $k$-th level.
The implementation of BGFC method is available as an open-source Julia package, \texttt{JuAC.jl} \cite{gitJuAC}. The mesh generator is developed from \texttt{Tetgen}\footnote{\cite{si2015tetgen}, a \texttt{C++} program for generating good quality tetrahedral meshes.}, which can adapt to the underlying crystalline structure and the a/c interface. 

Compared to the mesh generation and adaptation in two dimensions developed in our previous works \cite{liao2018posteriori, wang2018posteriori}, the main difficulties for complex crystalline defects in three dimensions are threefold: (i) The atomistic region is not guaranteed to be convex any more and the surface mesh needs to be constructed; (ii) A smooth transition region is required, especially for complex defects with large distortions, e.g., the (001)[100] edge dislocation in tungsten considered in this paper; (iii) A robust and efficient mesh adaptation is essential in three dimensions.
While this is an important subject, it is not the main focus of this paper, we leave the detailed constructions and discussions in a separate work \cite{CPCmesh}. 
We demonstrate the mesh generation algorithm with BCC crystalline structure in this paper, it can be extended to other crystalline structures such as FCC or HCP directly with minor modifications.

\subsection{Oneway multigrid strategy}
\label{sec:mstra}

In this section, we introduce the multigrid strategy for the geometry optimization of large-scale molecular mechanics. For simplicity, we focus on the oneway multigrid strategy also used in~\cite[Algorithm 3]{chen2017efficient}.

As discussed in~\cite{chen2017efficient}, compared to the oneway multigrid strategy, the full multigrid (FMG) strategy is even computationally more expensive than the {\it brute-force} optimization. Hence, we stick with the oneway multigrid strategy in the current work to make a direct comparison with the results in~\cite{chen2017efficient}. The extension to the FMG strategy with BGFC method may involve substantial additional technicality which would be included in a separate work. We will give a brief discussion in \S~\ref{sec:conclusion}.

Given the number of levels $L \in \N$, we first consider a sequence of tetrahedral mesh $\{\T_0, \T_1, \ldots, \T_{L}\}$ of $\Omega$.
Let $\T_{N}$ be the {\it canonical} tetrahedral mesh induced by $\L^{\Omega}$. For $T \in \T_k$, $k=1, \ldots, L$, we define the diameter of $T$, diam$(T):=\sup \{|x-y|, x,y \in T\}$. 
Motivated by the standard multigrid method, we denote $I_{i}^{j}$ as the interpolation operator (e.g., scattered interpolation) from level $i$ to level $j$ if $i<j$ and as the restriction operator if $i>j$. The restriction operator will not be used in practice due to the oneway structure. Algorithm~\ref{alg:oneway} is summarized as follows.


\begin{algorithm}[H]
\caption{Oneway multigrid strategy}
\label{alg:oneway}
\begin{enumerate}
	\item  Relax the coarse-grid problem $\mathbf{P}_0^{\rm QA}$ ($\mathbf{P}_0^{\rm BGFC}$) on the initial mesh $\T_0$ to obtain $u_0$ with a trivial initial guess.
	\item  For $k=1, \cdots, L$, relax the coarse-grid problem $\mathbf{P}_k^{\rm QA}$ ($\mathbf{P}_k^{\rm BGFC}$) on $\T_k$ to obtain $u_k$ with the initial guess $I^k_{k-1}u_{k-1}$.
	\item Solve the atomistic problem $\mathbf{P}^{\rm A}_N$ \eqref{eq::a} until convergence with the initial guess $I^{N}_{L}u_{L}$.
\end{enumerate}
\end{algorithm}


\subsection{Oneway adaptive multigrid with BGFC coarse problems (OAM-BGFC)}
\label{sec:algo}

In this section, we propose the oneway adaptive multigrid strategy with BGFC method (OAM-BGFC, Algorithm \ref{alg:bgfc}), which is the main algorithm in this paper. To make a comparison, the oneway adaptive multigrid strategy with QA approximation (OAM-QA, Algorithm \ref{alg:qa}) is also given in the~\ref{sec:appendix}. 

As discussed in \S~\ref{sec:bgfcm}, we use the heuristic gradient-based {\it a posteriori} error estimator. The local error estimator for BGFC method on each coarse-grid level, is then simply chosen as $\rho_T=\|\nabla u_k\|_{L^{2}(T)}$, where $T \in \T_k$ and $u_k$ is the solution of  coarse-grid problems $\mathbf{P}^{\rm QA}_k$ or $\mathbf{P}^{\rm BGFC}_k$ on the $k$-th level. A rigorous {\it a posteriori} error analysis for the BGFC method will be investigated in our future work.

Before introducing the adaptive multigrid algorithm, we first present the following mesh refinement strategy with the well-known D\"{o}rfler strategy \cite{Dorfler:1996}, where the atomistic/continuum partitions and the local mesh refinement in the continuum region are constructed on-the-fly on each coarse level.

\begin{algorithm}[H]
\caption{Mesh refinement for BGFC method.}
\label{alg:refine} 
Prescribe $0<\tau_1, \tau_2 <1$.  
\begin{enumerate}

    \item Given a partition $\T$ and the approximate solution $u$, compute the local error estimator $\rho_T = \|\nabla u\|_{L^{2}(T)}$ on each element $T \in \T$.

	\item Choose a minimal subset $\mathcal{M}\subset \T$ such that
	\begin{eqnarray}\label{eq:do}
		\sum_{T\in\mathcal{M}}\rho_{T}\geq\tau_1\sum_{T\in\T}\rho_{T}.
	\end{eqnarray}

	\item We can find the interface elements within $p$ layers of lattice spacing,  $\mathcal{M}_p:=\{T\in\mathcal{M}\bigcap(\T^{\b} \cup \T^{\c}): \textrm{dist}(T, \L^{\a})\leq p\}$. Choose $P>1$, find the first $p\leq P$ such that
	\begin{equation*}
		\sum_{T\in \mathcal{M}_p}\rho_{T}\geq \tau_2\sum_{T\in\mathcal{M}}\rho_{T},
		\label{eq:interface1}
	\end{equation*}
    let $\mathcal{M} = \mathcal{M}\setminus \mathcal{M}_p$. 
    \item Expand the atomistic region $\L^\a$ and the blending region $\Lambda^\b$ outward by $[\frac{p}{2}]$ and $p-[\frac{p}{2}]$ layers respectively. Bisect all elements $T\in \mathcal{M}$ to obtain new triangulation $\T$. 
\end{enumerate}
\end{algorithm}

The algorithm above gives a guidance of how to move the atomistic/continuum interface and refine the tetrahedral mesh in the continuum region based on the local error estimators, which follows from the two dimensional setup in \cite[Algorithm 3]{wang2018posteriori} and is adapted to three dimensional problems. 
It has been effectively implemented in our mesh generator. 

During the adaptive mesh refinement procedure summarized in Algorithm~\ref{alg:refine}, we first apply the well-known D\"{o}rfler strategy \eqref{eq:do}, which is a widely used marking strategy to enforce error reduction, to choose a subset $\mathcal{M}$ for model adjustment and mesh refinement. The atomistic and blending regions are then equally expanded according to $\mathcal{M}_p$, which is a set of the marked elements near the atomistic/continuum interface. We apply the techniques developed in~\cite{CPCmesh} to bisect all marked elements in the continuum region. Once all of these steps are accomplished, the new mesh on the next coarse-grid level is constructed. The adaptive parameters in Algorithm~\ref{alg:refine} are fixed as $\tau_1=0.3$ and $\tau_2=0.5$ throughout all the numerical experiments conducted in \S~\ref{sec:ne}.

We admit that the proposed model refinement strategy seems to be ad hoc since we exploit the {\it a priori} knowledge that the radius $R^{\a}$ of the atomistic region and the width $L^{\b}$ of the blending region are kept identical (cf.~\S~\ref{sec:bgfcm}). However, the main purpose of this paper is to develop a multigrid strategy for the geometry optimization by utilizing the idea of adaptivity. A more general algorithm requires a carefully designed strategy for model refinement, which may need to combine the ideas such as the residual-based error estimator based on the stress formulation \cite{wang2018posteriori, wang2021posteriori}, and will be investigated in our future work.

Finally, based on Algorithm \ref{alg:refine}, we give the following adaptive multigrid strategy based on the $\mathbf{P}^{\rm BGFC}_k$ coarse problem (OAM-BGFC).

\begin{algorithm}[H]
\caption{Oneway adaptive multigrid strategy with BGFC coarse problem (OAM-BGFC).}
\label{alg:bgfc}
\begin{enumerate}
	\item Prescribe the optimization step $\mu_k$ and the tolerance ${\rm tol}_k$ on $k$-th level.
	
	\item Relax the BGFC problem $\mathbf{P}^{\rm BGFC}_0$ defined by \eqref{eq::pbgfc} on the initial mesh $\T_0$ until reaching the tolerance ${\rm tol}_0$ or exceeding $\mu_0$ times to obtain $u_0$ with a trivial initial guess (for example $\hat{u}_0$), compute $n_0$, set $k=1$.
	
	\item Carry out the mesh refinement based on Algorithm \ref{alg:refine} to obtain $\T_{k}$, solve the $\mathbf{P}^{\rm BGFC}_k$ defined by \eqref{eq::pbgfc} with the initial guess interpolated from the solution on $\T_k$ until reaching the tolerance ${\rm tol}_k$ or exceeding $\mu_k$ times, compute $n_k$, if $n_k \leq 0.1 N$, $k=k+1$, goto Step 3 again; Otherwise, goto Step 4.
	\item Solve the atomistic problem $\mathbf{P}^{\rm A}_N$ defined by \eqref{eq::a} until convergence with the initial guess interpolated from the solution on $\T_{k}$.
   
	\end{enumerate}
\end{algorithm}

It is worthwhile mentioning that in the stopping criteria of the BGFC coarse problems in Algorithm \ref{alg:bgfc}, the values of parameters $\mu_k$ and ${\rm tol}_k$ heavily influence the performance of the resulting adaptive multigrid strategy. Hence, we will provide a detailed study of these parameters in the following section (cf.~\S~\ref{sec:ps}).

\section{Numerical Experiments}
\label{sec:ne}

In this section, we implement and test the main algorithm (Algorithm \ref{alg:bgfc}) for three prototypical examples of three dimensional localised defects: single vacancy, micro-crack and edge dislocation. 
We use tungsten (W) in all numerical experiments, which has a body-centered cubic (BCC) crystal structure in the solid state. The EAM potential \cite{Daw1984a} is used to model the interatomic interaction. The cut-off radius is chosen as $r_{\rm cut} = 5.6$\AA, which includes up to the third neighbour interaction. All numerical experiments are tested with the clamped boundary conditions applied in three directions. To avoid the boundary effects, we introduce several layers of ghost atoms whose thickness are greater than $r_{\rm cut}$ outside of the domain of interest.

For BGFC method on each coarse level, replicating the setting in \cite{colz2016}, we implement the equivalent ghost force removal formulation \eqref{eq:gfc} instead of the ``renormalization formulation" \eqref{eqn:Eqc}. As discussed in Remark~\ref{re:disloc}, for the sake of simplicity, we choose $\hat{u}_0=0$ in~\eqref{eq:gfc} for all types of crystalline defects considered in this paper. The blending function $\beta$ is obtained in a preprocessing step by approximately minimizing $\|\nabla^2 \beta\|_{L^2}$, as described in detail in~\cite{luskin2013formulation}. 

For the the choice of the optimization solver, according to \cite{chen2017efficient}, \texttt{CG\_DESCENT} by Hager and Zhang \cite{hager2006algorithm} gives a better compromise between accuracy and efficiency compared with \texttt{L-BFGS} \cite{liu1989limited}. Moreover, the line search implementation in \texttt{CG\_DESCENT} appears to be the most robust on coarse levels. Hence, throughout this paper, the coarse-grid problems \eqref{eq::uka} and \eqref{eq::pbgfc} are both solved by \texttt{CG\_DESCENT}, which is implemented by an open source \texttt{Julia} package, \texttt{Optim.jl} \cite{mogensen2018optim}. For comparison, we also use \texttt{CG\_DESCENT} to solve the atomistic problem~\eqref{eq::a}.
We denote $u^{{\rm qa, (\mu_k)}}_k$ and $u^{{\rm bgfc, (\mu_k)}}_k$ as their corresponding approximated solutions after $\mu_k$ steps of \texttt{CG\_DESCENT}.

We apply the oneway adaptive multigrid strategies (Algorithm \ref{alg:bgfc} and Algorithm \ref{alg:qa}) for three defect configurations: single vacancy, micro-crack and [100](001) edge dislocation in three dimensional tungsten BCC lattice. Let $\textbf{g}:=\nabla \E(u)$, we use the same stopping criteria $|\textbf{g}|_{\infty} < {\rm tol} = 10^{-4}$ for single vacancy and micro-crack while $|\textbf{g}|_{\infty} < {\rm tol} = 10^{-3}$ for edge dislocation. To make a comparison so that the efficiency of the main algorithm (Algorithm \ref{alg:bgfc}) can be clearly shown, we additionally implement and test several algorithms including the {\it brute-force} optimization, the oneway multigrid strategies with the QA approximation given in \cite{chen2017efficient} and the adaptive QA approximation (cf. Algorithm \ref{alg:qa}).
All tests are performed on a Linux cluster with {\tt AMD EPYC-Rome Processor} with 96 cores and 2TB memory.

\subsection{Parameter study}
\label{sec:ps}

As discussed briefly in \S~\ref{sec:algo}, the parameters $\mu_k$, ${\rm tol}_k$ and $n_k$ in the stopping criteria in Algorithm \ref{alg:bgfc} heavily influence the efficiency of the resulting oneway adaptive multigrid strategy. In this section, we take the single vacancy as the benchmark example to study the parameters in Algorithm \ref{alg:bgfc}, with the following setup. 

\begin{figure}[htb]
	\centering 
	\includegraphics[height=6cm]{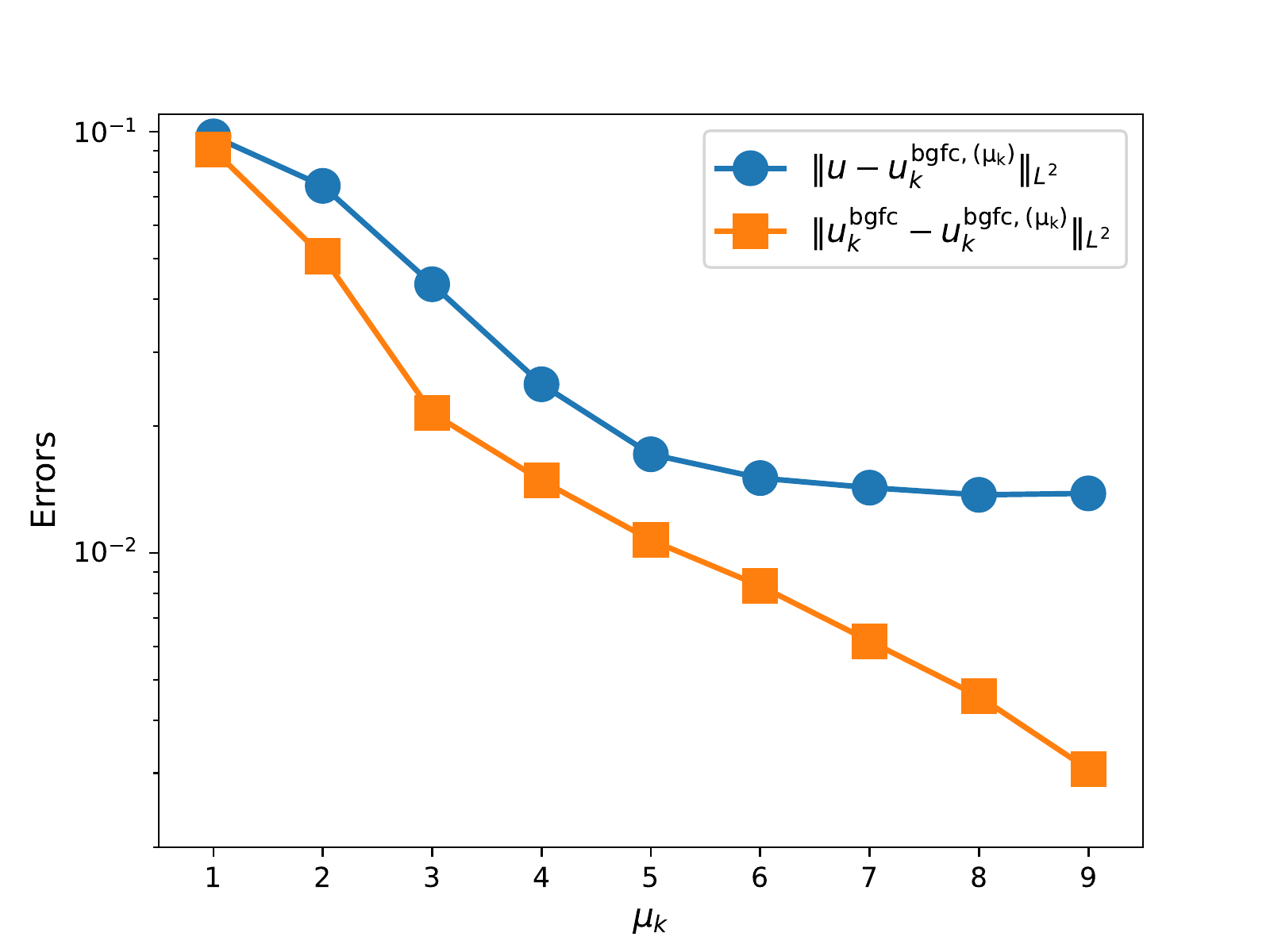}
 	\caption{Total error $\| u - u_k^{\rm bgfc, (\mu_k)}\|_{L^2}$ and algebraic error $\|u_k^{\rm bgfc} - u_k^{\rm bgfc, (\mu_k)}\|_{L^2}$ with respect to the iteration number $\mu_k$.}
	\label{fig:algvstot}
\end{figure}

We first set $R^{\rm a}$, $L^{\rm b}$ and $ R_{\Omega}$ as $4r_0$, $4r_0$, and $30r_0$ respectively, where $r_0$ is the lattice constant of W. We compute and draw the total error $\| u - u_k^{\rm bgfc, (\mu_k)}\|_{L^2}$ and the algebraic error $\|u_k^{\rm bgfc} - u_k^{\rm bgfc, (\mu_k)}\|_{L^2}$ with respect to $\mu_k$, where $u$ and $u_k^{\rm bgfc}$ are the solutions to \eqref{eq::a} and \eqref{eq::pbgfc} respectively, while $u_k^{\rm bgfc, (\mu_k)}$ is the approximated solution to \eqref{eq::pbgfc} after $\mu_k$ steps of optimization. 
Figure \ref{fig:algvstot} shows that the algebraic error $\|u_k^{\rm bgfc} - u_k^{\rm bgfc, (\mu_k)}\|_{L^2}$ decays as $\mu_k$ increases. However, the total error $\|u - u_k^{\rm bgfc, (\mu_k)}\|_{L^2}$ is levelling out when $\mu_k\geq 6$. To explain this, it is straightforward to see that 
\[
\| u - u_k^{\rm bgfc, (\mu_k)}\|_{L^2} \leq \|u - u_k^{\rm bgfc}\|_{L^2} + \|u_k^{\rm bgfc} - u_k^{\rm bgfc, (\mu_k)}\|_{L^2}.
\]
The total error $\|u - u_k^{\rm bgfc, (\mu_k)}\|_{L^2}$ is bounded by the modeling error $\|u - u_k^{\rm bgfc}\|_{L^2}$ when $\mu_k$ is large. This observation motivates us that solving the coarse-grid problem on each coarse-grid level with a very high accuracy may gild the lily since the dominant part is always the modeling error, which also verifies the similar argument in \cite{chen2017efficient}, where the QA approximation is employed as the coarse-grid model. 

We record the CPU times (in seconds) of the oneway adaptive multigrid strategy with BGFC method (OAM-BGFC) by applying several stopping criterion on each coarse-grid level, including fixed accuracy ${\rm tol}_k$, fixed the number of optimization iteration $\mu_k$, and the {\it accuracy control} (flexible ${\rm tol}_k$ dependent of $k$) on each level. Given the initial tolerance ${\rm tol}_0=10^{-1}$, the accuracy control on each coarse-grid level is then enforced by simply setting ${\rm tol}_k = 2^{-k} \cdot {\rm tol}_0$. For comparison, the CPU time of the {\it brute-force} optimization is also tested and listed in Table \ref{tab:para}. We consider the reference atomistic problem \eqref{eq::a} with {\it dofs} $N=4.32\times 10^5$ and $N=3.456\times 10^6$.

We observe from Table \ref{tab:para} that the accuracy control strategy is the most efficient stopping criteria. Hence, we will use this strategy as the stopping criterion in Algorithm \ref{alg:bgfc} throughout the work without any tuning. 
We also observe that, fixed iteration number strategies ($\mu_k = 3, 8$) are more efficient than the fixed accuracy strategies (${\rm tol}_k = 10^{-2}, 10^{-3}$). As a matter of fact, tolerance ${\rm tol}_k$ characterize the algebraic error $\|u_k^{\rm bgfc} - u_k^{\rm bgfc, (\mu_k)}\|_{L^2}$ on each coarse level, therefore solving the BGFC method with sufficiently small ${\rm tol}_k$ will not help improve the efficiency due to the dominant modeling error. 

\begin{table}[htbp]
\label{tab:para}
\centering
\begin{tabular}{l|ll}
\hline
Stopping criteria vs Time (s) & CPU time (N = $4.32\times 10^5$) & CPU time (N = 3.456$\times 10^6$) \\ \hline
fixed accuracy 1e-2  &  2847 & 20242 \\ 
fixed accuracy 1e-3  &  2321 & 14888 \\ 
fixed number 3 & 2593 & 15847 \\ 
fixed number 8 & 2065 & 13418 \\ 
accuracy control & \textbf{1832} & \textbf{9275}  \\ 
brute-force & 2215 & 21842 \\ 
\hline
\end{tabular}
\caption{CPU time vs. Stopping criteria, bold font indicates best performance}
\end{table}

We also note that the cost of mesh generation and adaptation is negligible compared with that of the optimization for BGFC method on each coarse level. The proportion of the CPU time for generating and refining the mesh are less than 1\% in total CPU time for all numerical experiments shown in Section \ref{sec:ne}, thanks to the efficiency of our mesh generator \cite{CPCmesh}.

\subsection{Single vacancy}
\label{sec:sv}

We first consider the case with one single vacancy at the origin. To create the vacancy for tungsten, we simply remove the atom at the origin. 
The reference atomistic systems have $30^3, 60^3, 120^3, 180^3, 300^3$ and $370^3$ unit cells such that the \textit{dofs} ($N$) of the reference atomistic problem ranges from $5.4\times 10^4$ to $1.014\times 10^8$. Clamped boundary conditions are imposed in all three directions with 3 layers of ghost atoms on each direction.

\begin{figure}[htb]
	\centering 
	\subfigure[Displacement $|u|$]{\label{fig:sing-disp}
	\includegraphics[height=5.5cm]{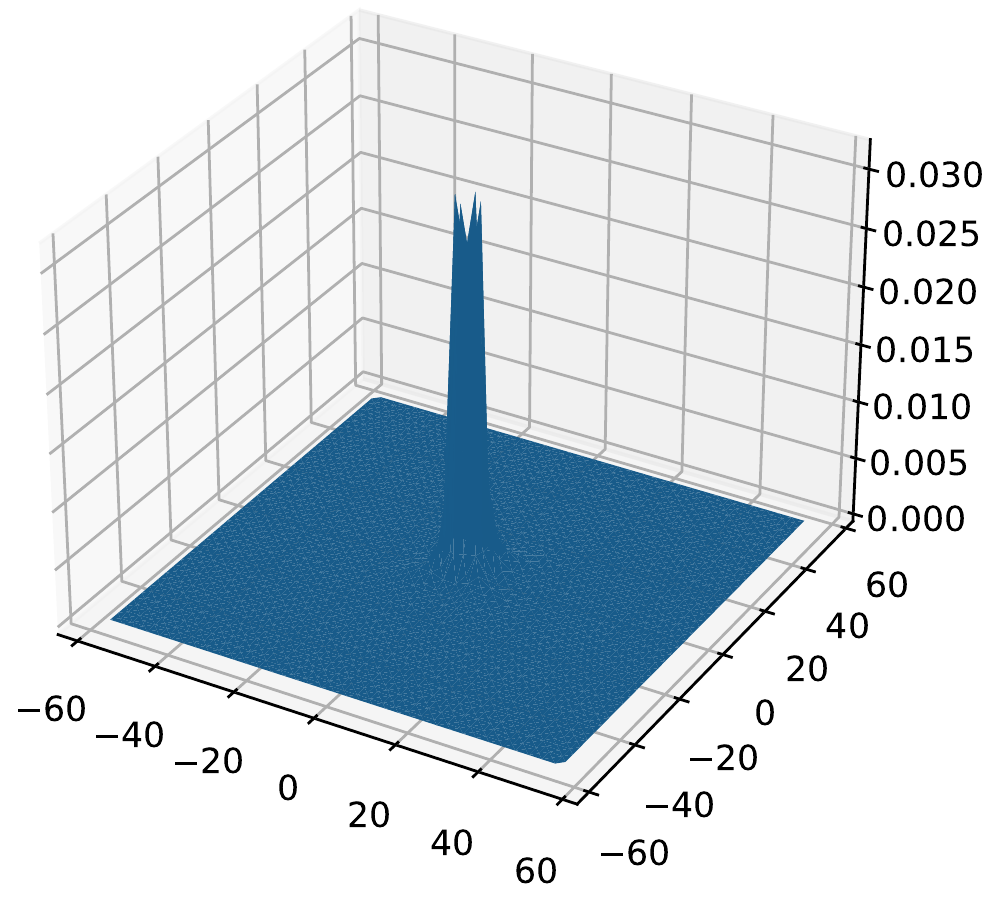}}
	\quad 
	\subfigure[Decay of $|u(\ell)|$]{\label{fig:sing-decay}
	\includegraphics[height=5.5cm]{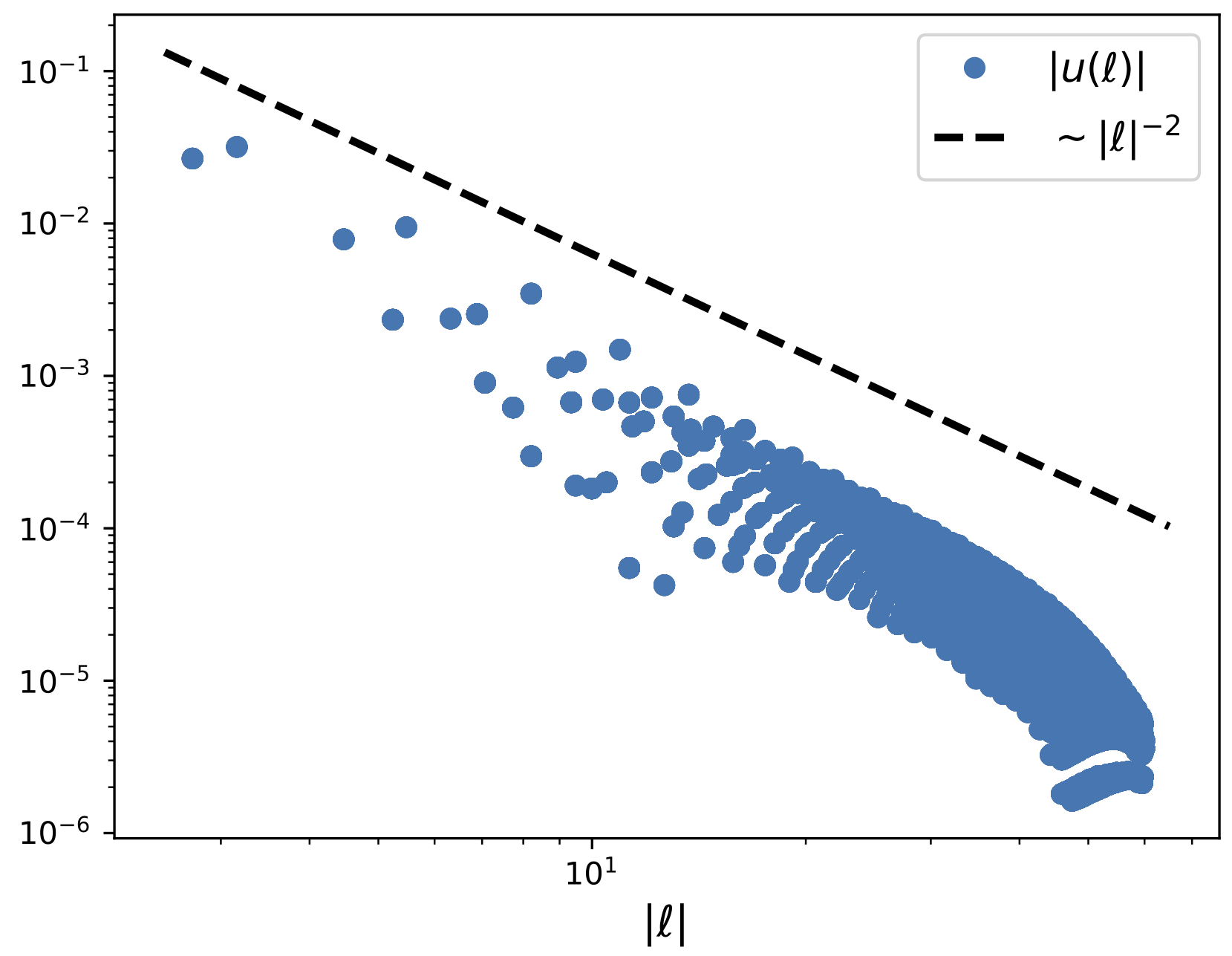}}
 	\caption{The displacement and the decay of the equilibrium for single vacancy.}
	\label{fig:sing}
\end{figure}

Figure \ref{fig:sing-disp} plots the displacement field over a centered slice in the $z$ direction of the $30^3$ system. It is verified by Figure \ref{fig:sing-decay} that the single vacancy is a highly-localized defect with $|\ell|^{-2}$ decay, that is, $|u(\ell)|\lesssim |\ell|^{-2}$, where $|\ell|$ is the distance to the defect core, consistent with the theoretical results in \cite[Theorem 1]{2013-defects}.

The parameters in the oneway adaptive multigrid with BGFC method (OAM-BGFC) are determined in Section \ref{sec:ps}. As discussed before, we also test the oneway adaptive multigrid method with QA approximation (OAM-QA, Algorithm \ref{alg:qa}) to make a comparison. We use fixed iteration number $\mu_k=3$ as the stopping criteria on each coarse level, which is the same as the setting in~\cite{chen2017efficient}. See also the~\ref{sec:appendix} for a detailed discussion. 

\begin{figure}[htb]
	\centering 
	\subfigure[CPU time]{
	\label{single-cpu-time}
	\includegraphics[height=5.5cm]{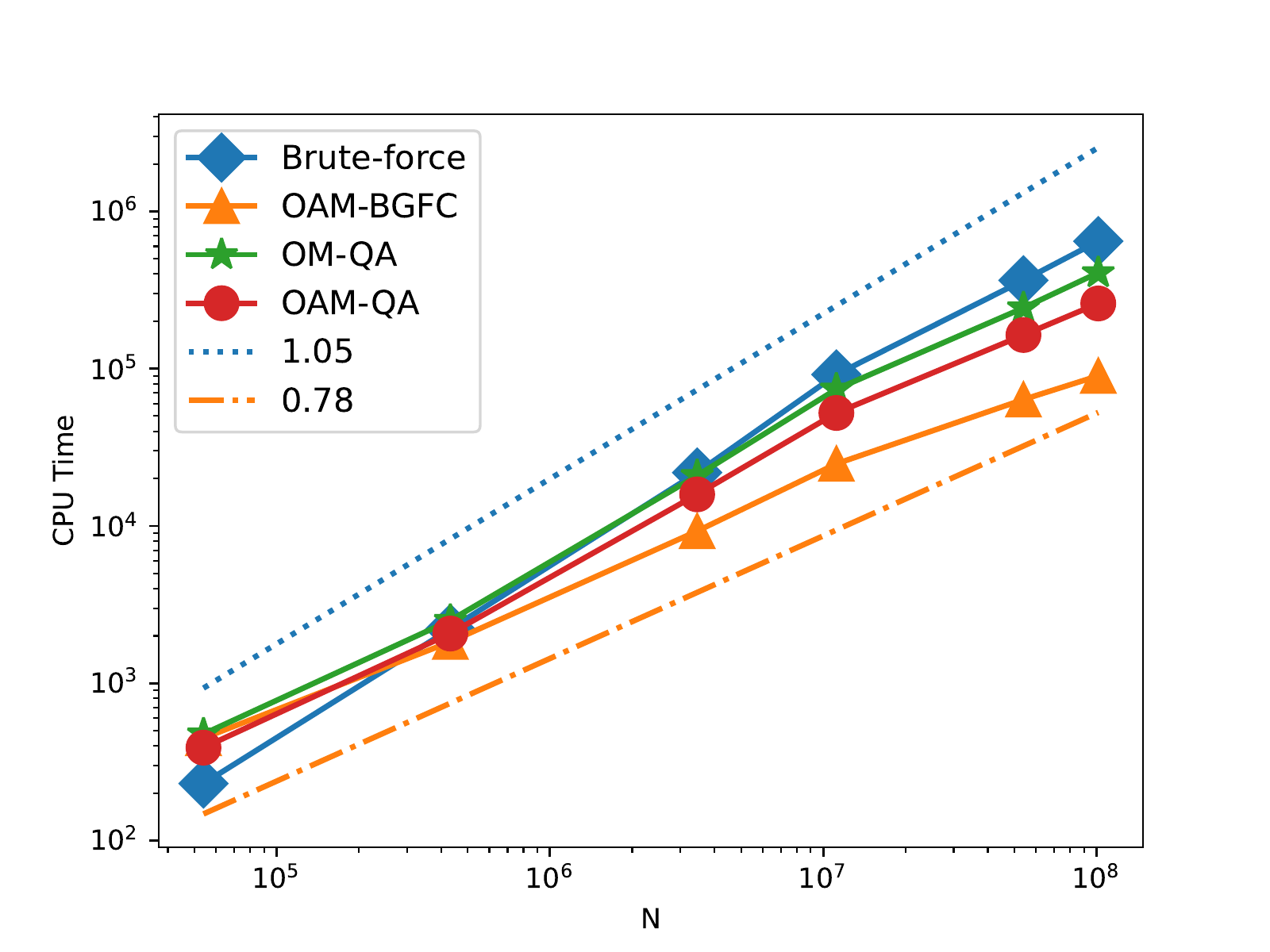}}
	\quad
	\subfigure[Time ratio]{
	\label{single-time-ratio}
	\includegraphics[height=5.5cm]{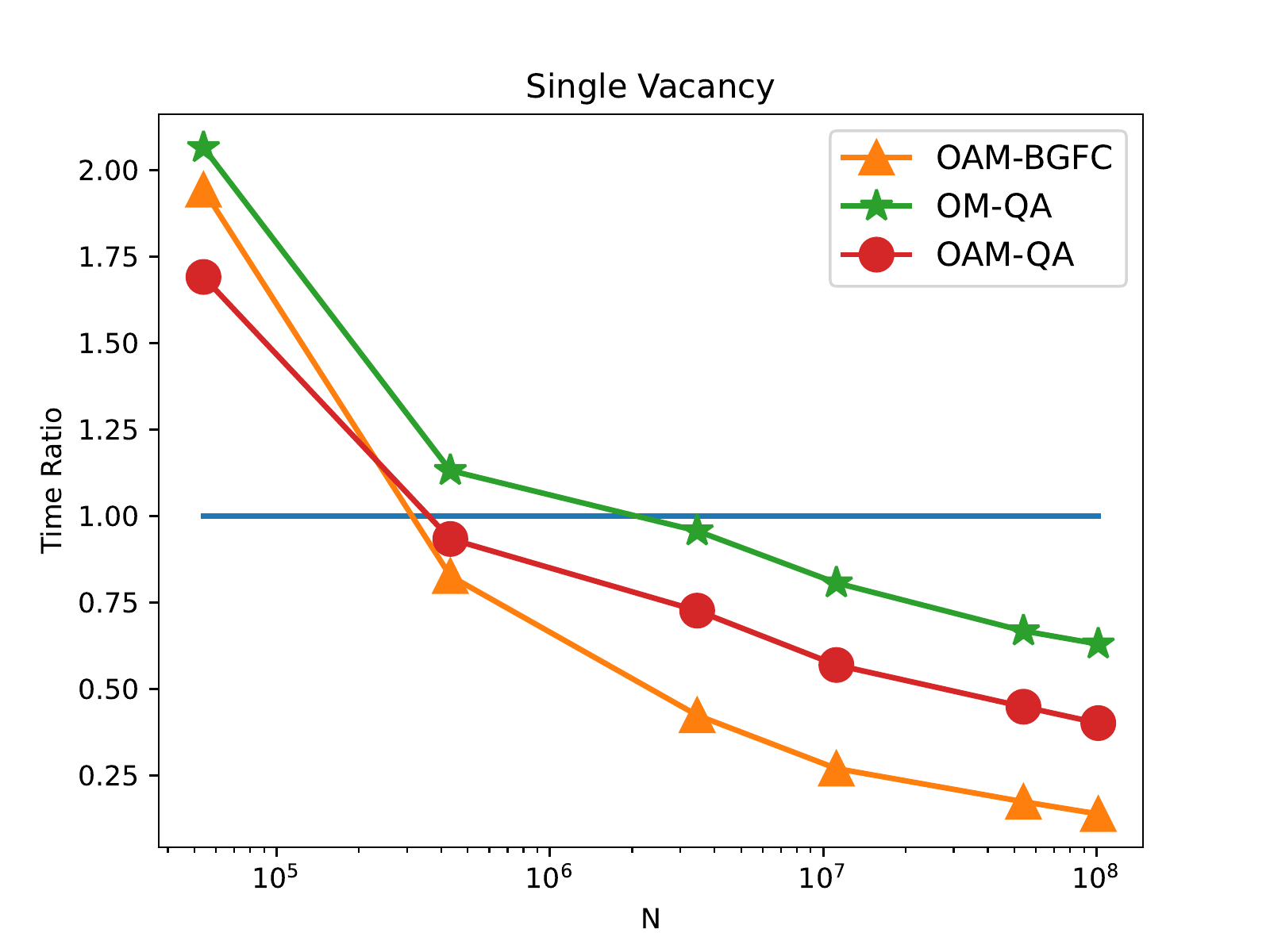}}
	\caption{CPU times of the oneway adaptive multigrid method with BGFC approximation and its comparision with the brute-force optimization and oneway multigrid with other types of coarse approximations, for single vacancy. }
	\label{fig:sing-time}
\end{figure}

Figure \ref{single-cpu-time} presents the CPU times of the {\it brute-force} optimization (blue line), the oneway multigrid method proposed in \cite{chen2017efficient} (OM-QA, green line), the oneway adaptive multigrid method with QA relaxation (OAM-QA, red line), and the oneway adaptive multigrid with BGFC method (OAM-BGFC, orange line) with respect to $N$ in the log–log scale. Figure \ref{single-time-ratio} shows the ratio of CPU times between these methods and the brute-force optimization. 
We observe that the OAM-BGFC method scales sublinearly ($O(N^{0.78}))$ while the {\it brute-force} optimization and other methods scale like $O(N^{1.05})$ asymptotically. More importantly, compared with the {\it brute-force} optimization, the saving of this scheme are around 80\% for systems with a hundred millions ($10^8$) atoms.

\begin{figure}[htb]
	\centering 
	\includegraphics[height=6cm]{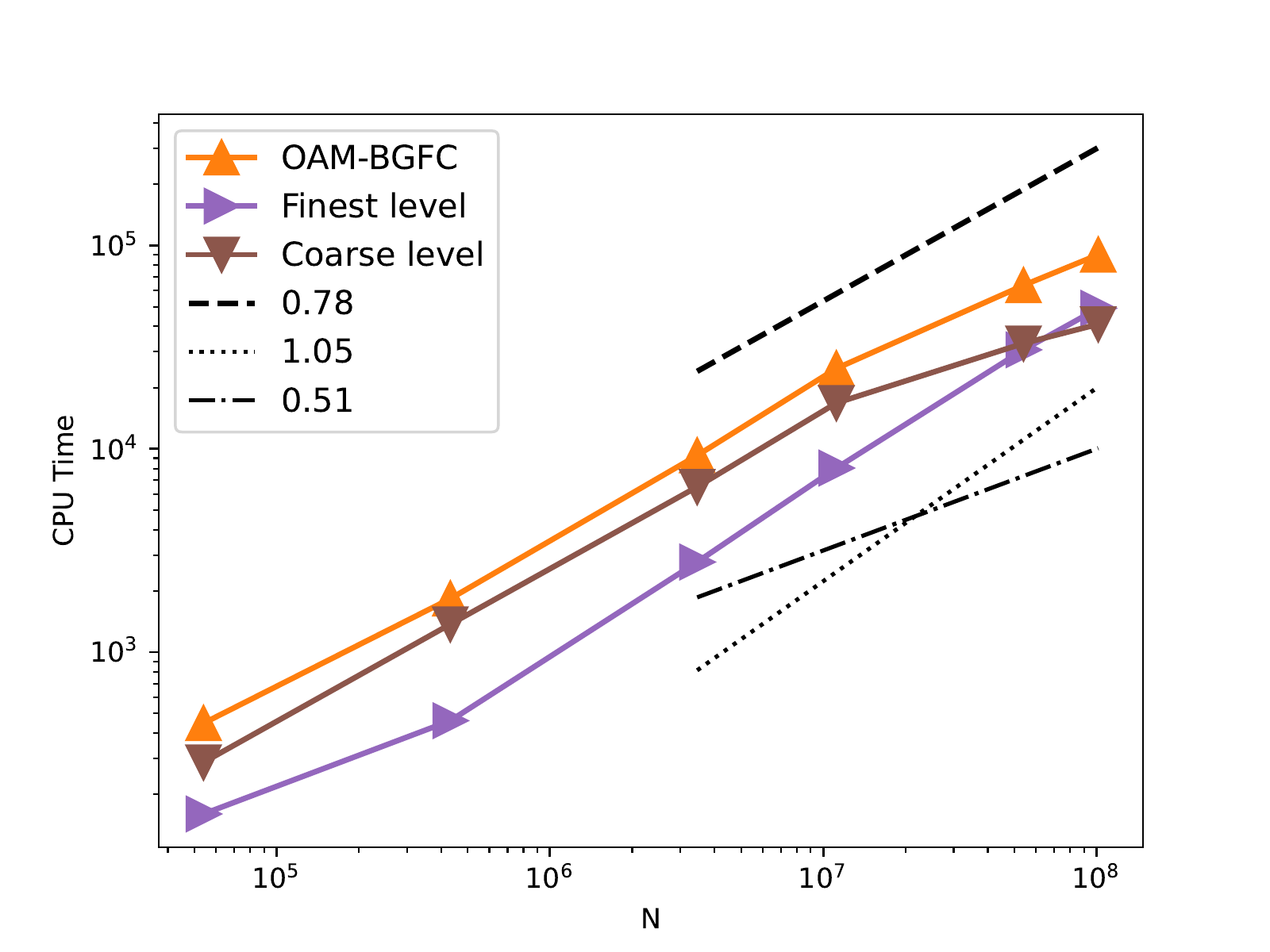}
	\caption{Time constitution of Algorithm \ref{alg:bgfc} for single vacancy.}
	\label{fig:septime}
\end{figure}

Figure \ref{fig:septime} demonstrates the time constitution to explain the sublinear scaling complexity of the OAM-BGFC method. We observe that the cost of the BGFC method on coarse levels scales sublinearly (brown line), although on the finest atomistic level, the computational cost is still linear (purple line). 

We note that the equilibrium of a single vacancy is localized with an $|\ell|^{-2}$ decay of the displacement filed, and a sequence of BGFC approximations provide a very ``good" initial guess for the finest atomistic optimization, which leads to the sublinear cost of the adaptive multigrid geometry optimization. We will show in \S~\ref{sec:ed} that, for more complex defects with longer-range elastic field such as the edge-dislocation (cf.~Figure~\ref{dislocation-decay}), the brute-force  geometry optimization incurs superlinear cost, and the adaptive multigrid strategy can reduce the complexity rate significantly in that case as well (cf.~Figure~\ref{fig:dislocation-time}).

\begin{figure}[H]
	\centering 
	\includegraphics[height=5.5cm]{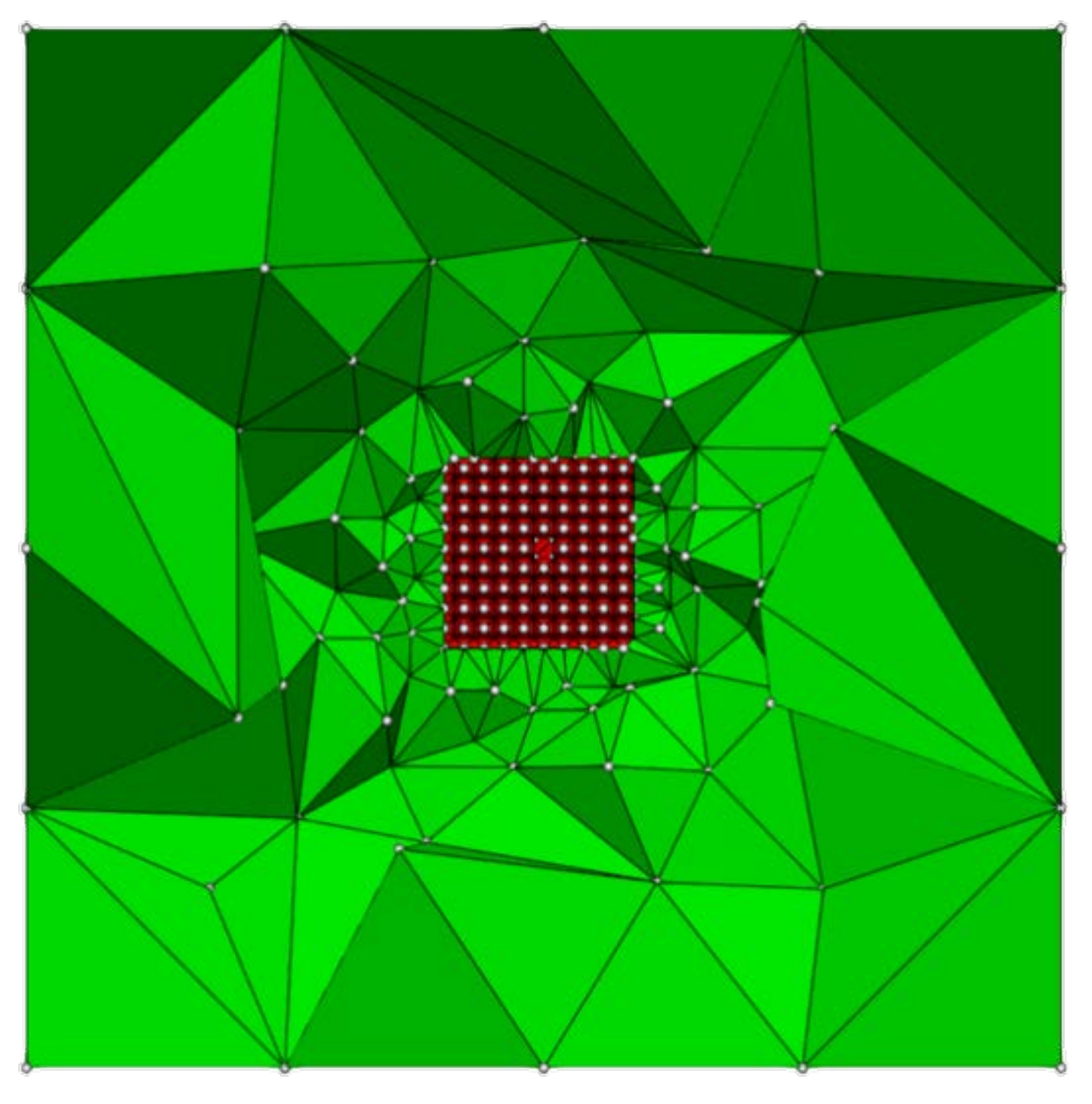}
	~
	\includegraphics[height=5.5cm]{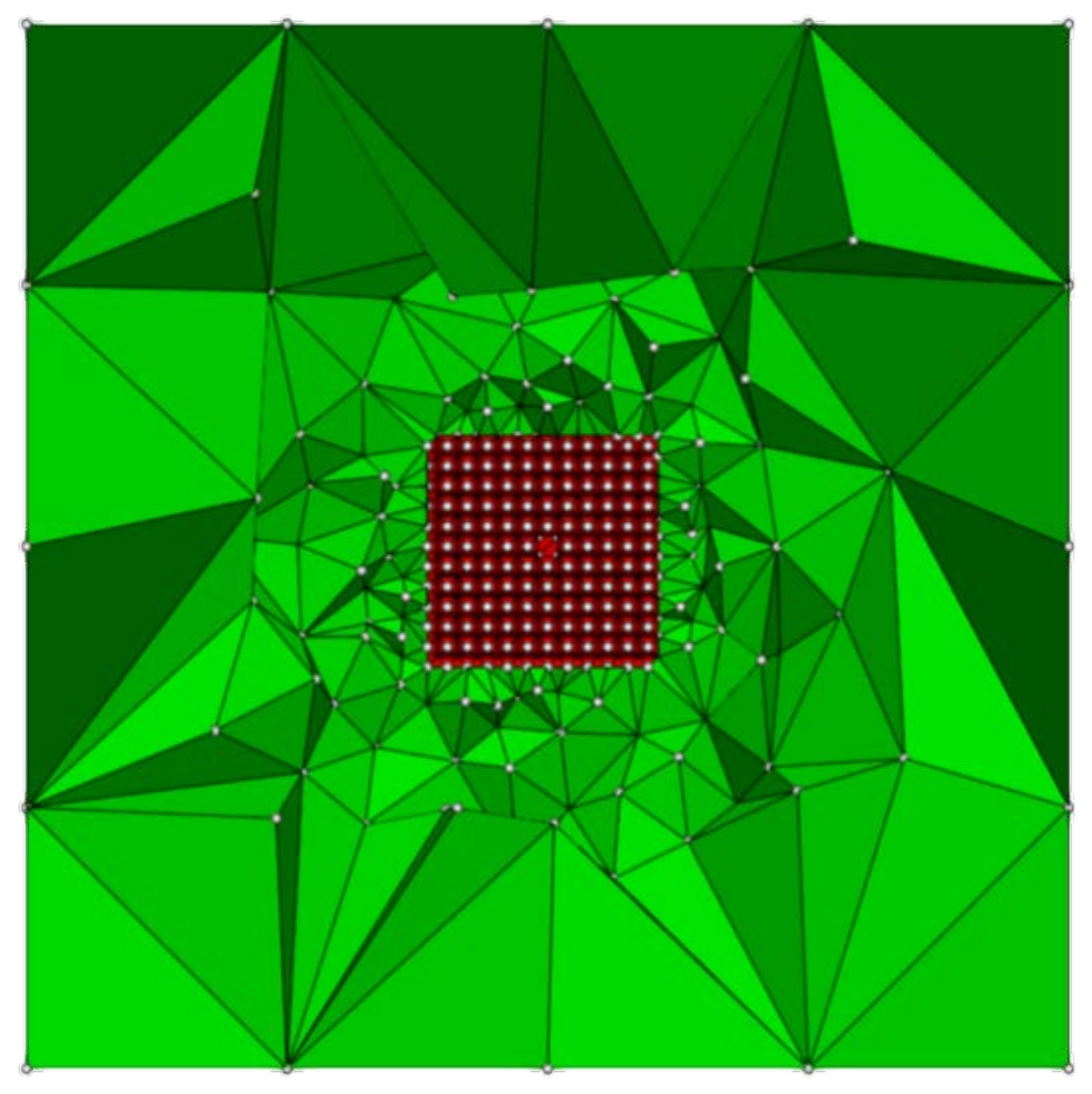}
	\\
	\includegraphics[height=5.5cm]{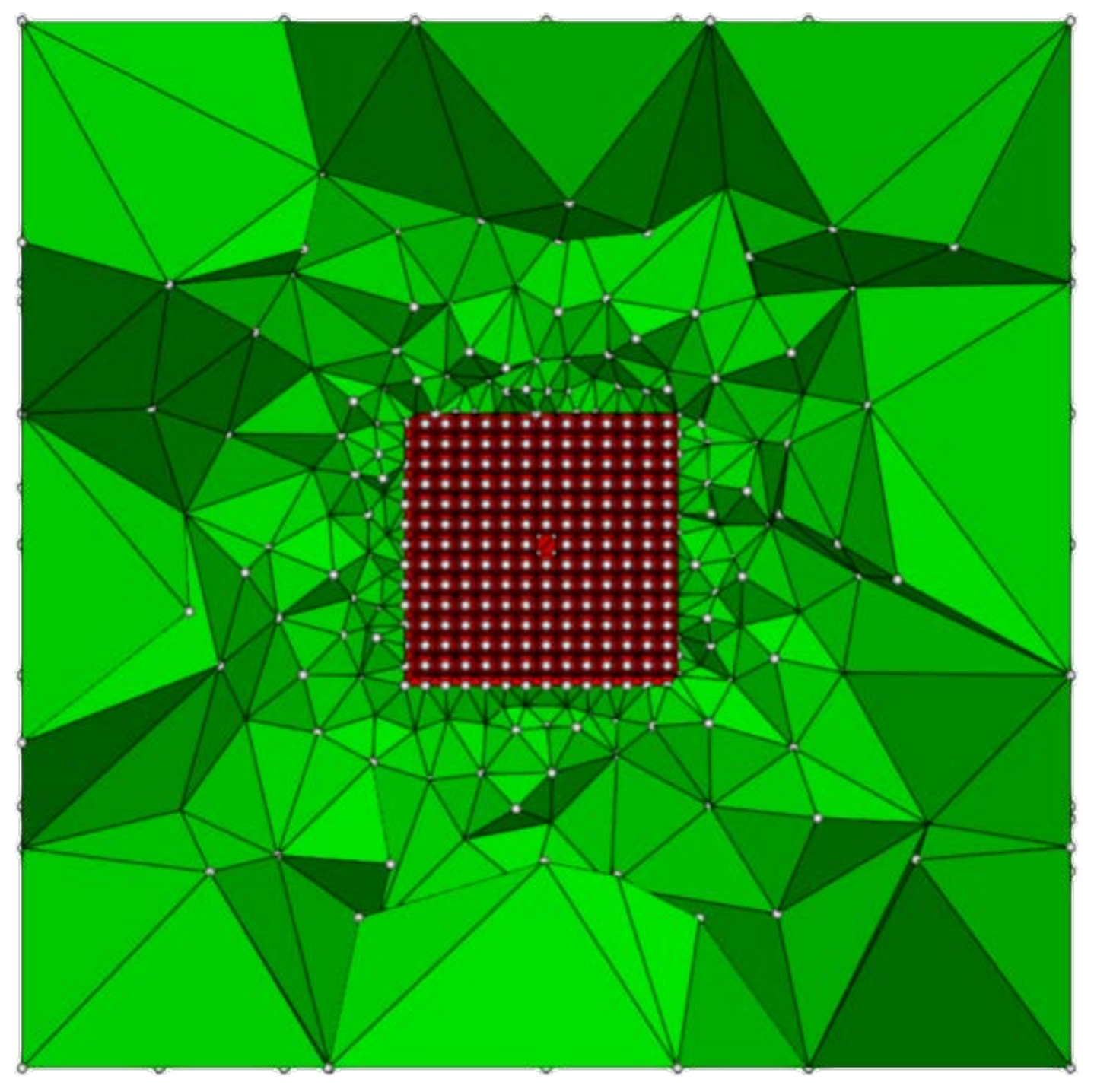}
	~
	\includegraphics[height=5.5cm]{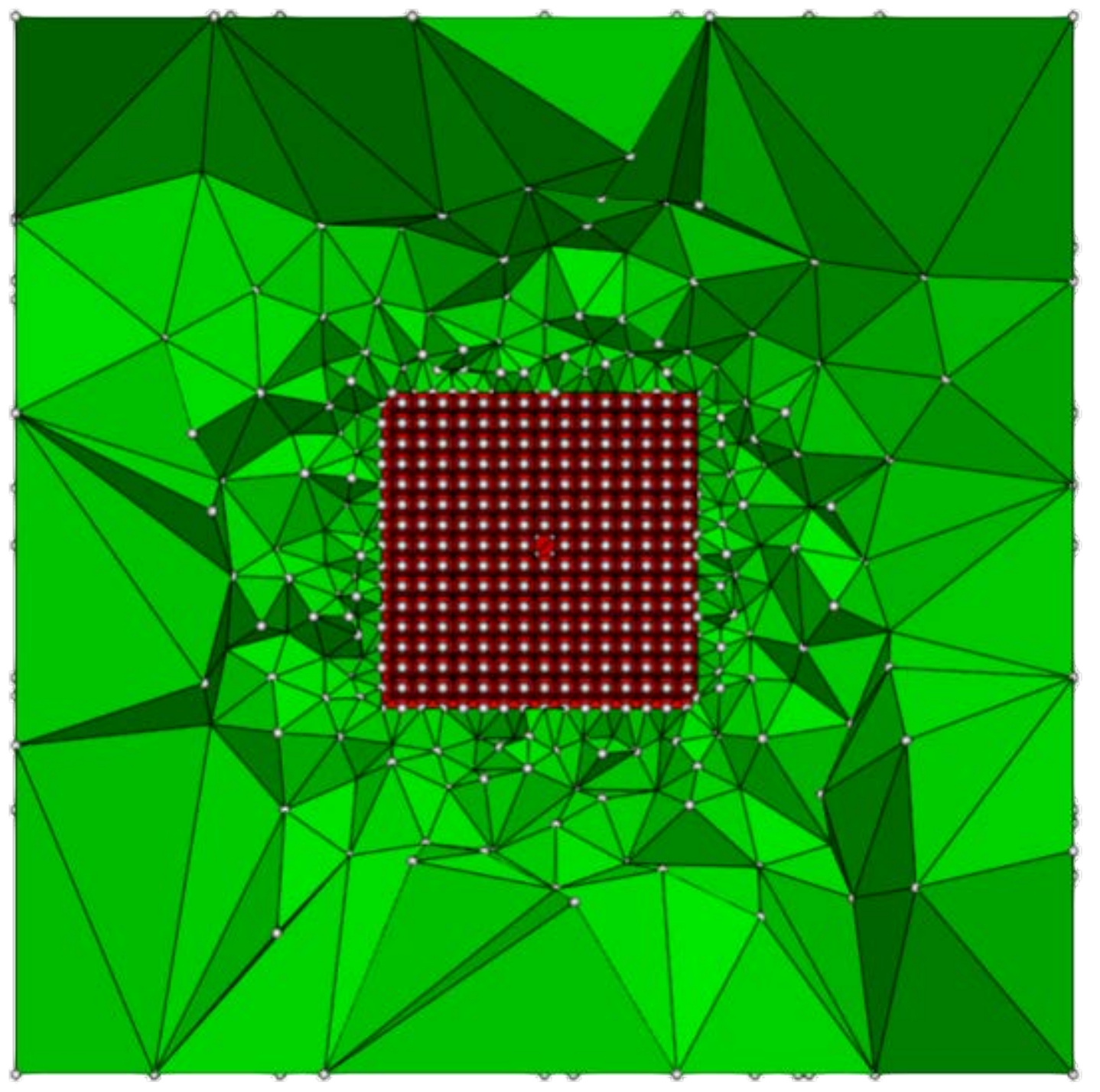}
 	\caption{Adaptive mesh refinement (degrees of freedom from upper left to lower right: 2142, 3880, 6275, 8051) of OAM-BGFC for a single vacancy.}
	\label{fig:sing-adpt-mesh-new-bgfc}
\end{figure}

Figure \ref{fig:sing-adpt-mesh-new-bgfc} shows the mesh evolution in the adaptive process for the OAM-BGFC for the $30^3$ system. The mesh evolution for the OAM-QA is also given by Figure \ref{fig:sing-adpt-mesh-new} in the \ref{sec:appendix}.

\subsection{Micro-crack}
\label{sec:mc}

Next, we will consider micro-crack. To create a micro-crack in the tungsten lattice, we remove seven adjacent atoms in the center of the x-y plane along the x direction. While this is not technically a “crack”, it serves as an example of a localised defect with an anisotropic shape.
We consider systems with $30^3, 60^3, 120^3, 180^3, 300^3$ and $370^3$ unit cells for the micro-crack case, which is the same as the single-vacancy case. Clamped boundary conditions are used in all three directions with three layers of ghost atoms for each direction. 

\begin{figure}[H]
	\centering 
	\subfigure[Displacement $|u|$]{\label{fig:crack-disp}
	\includegraphics[height=6cm]{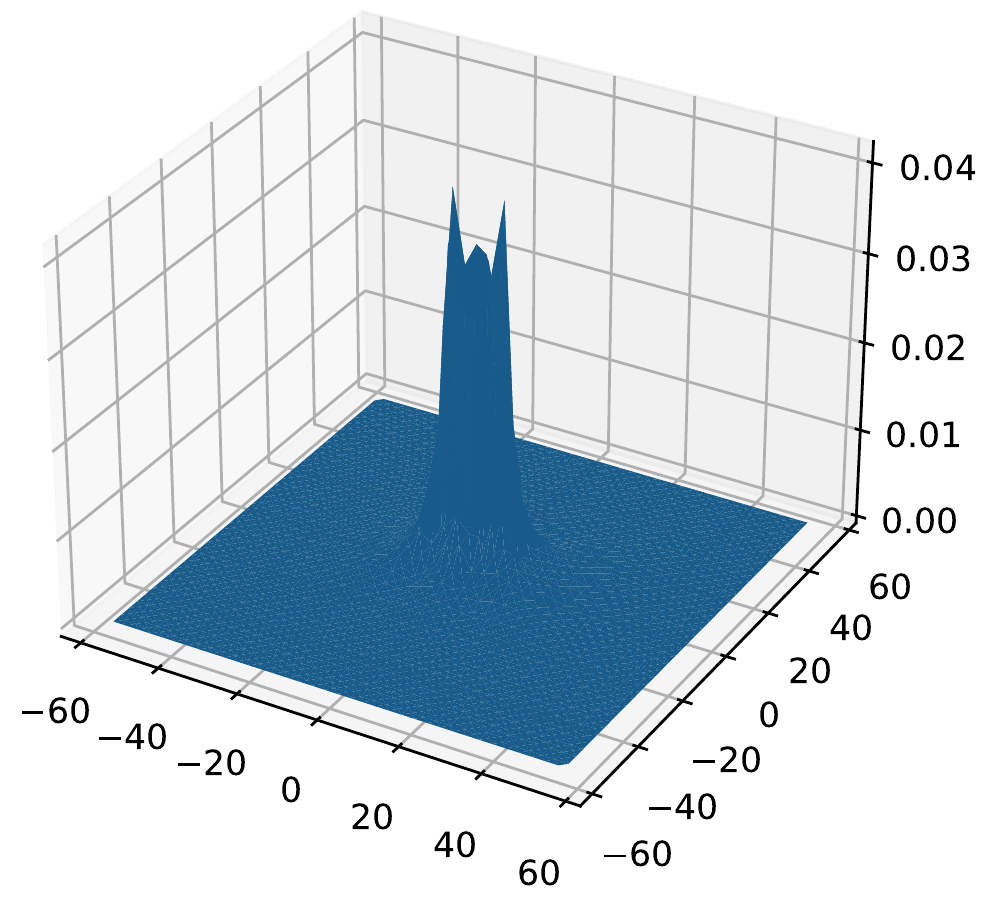}}
	\quad
	\subfigure[Decay of $|u(\ell)|$]{\label{fig:crack-decay}
	\includegraphics[height=6cm]{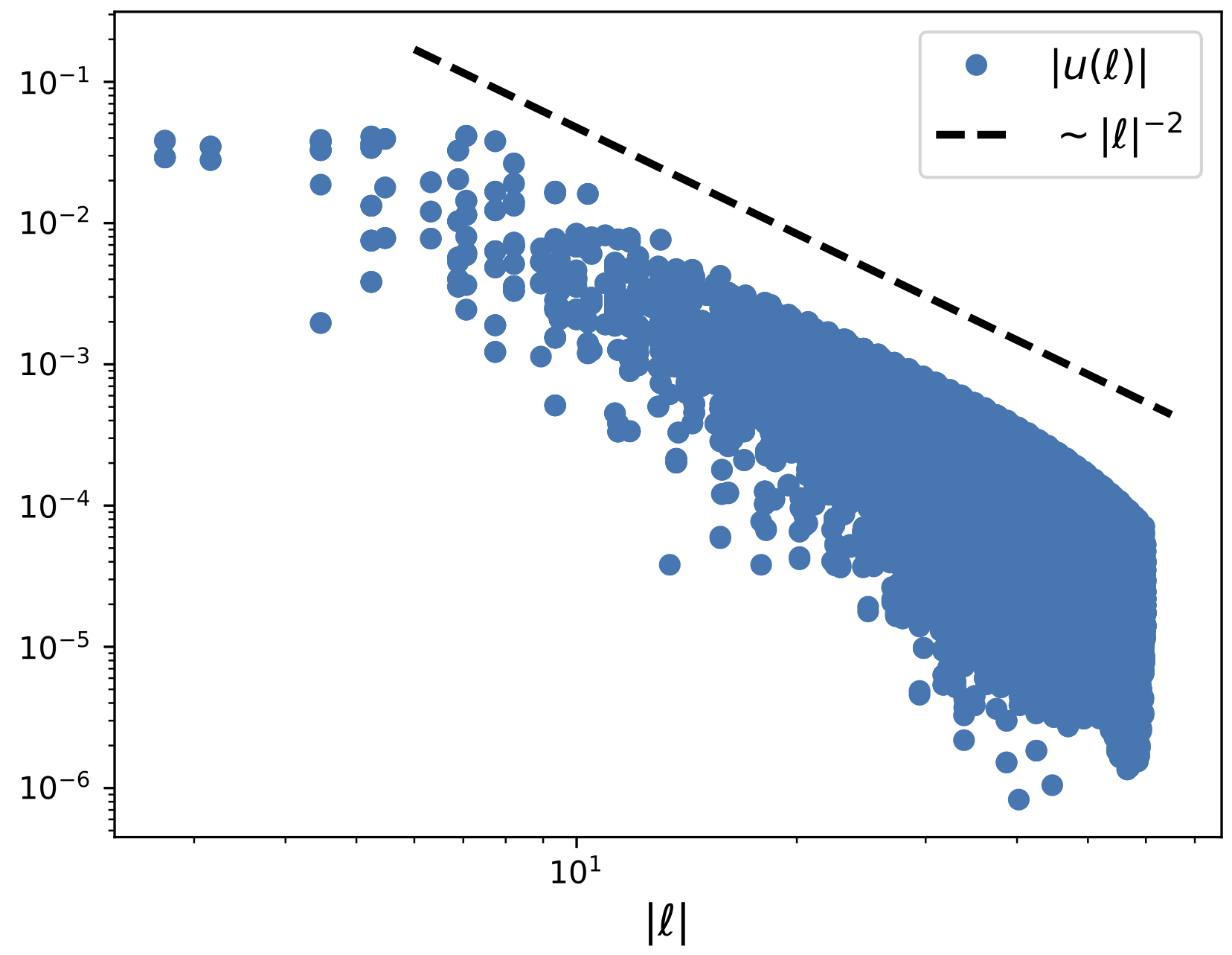}}
 	\caption{The equilibrium displacement and its decay for micro-crack.}
	\label{fig:crack}
\end{figure}

Figure \ref{fig:crack-disp} plots the displacement field over a centered slice in the $z$ direction of $30^3$ system and its decay with respect to the distance to the defect core is shown in Figure \ref{fig:crack-decay}. Around the micro-crack, one can see the anisotropic shape of the displacement field. Compared to Figure \ref{fig:sing-disp}, we observe that the interaction of micro-crack is stronger, though the far-field decay of the micro-crack is the same as that of the single vacancy. Therefore, solving the geometry optimization problem for micro-crack is relatively more difficult than that for single vacancy.

\begin{figure}[H]
	\centering 
	\subfigure[CPU time]{
	\label{crack-cpu-time}
	\includegraphics[height=6cm]{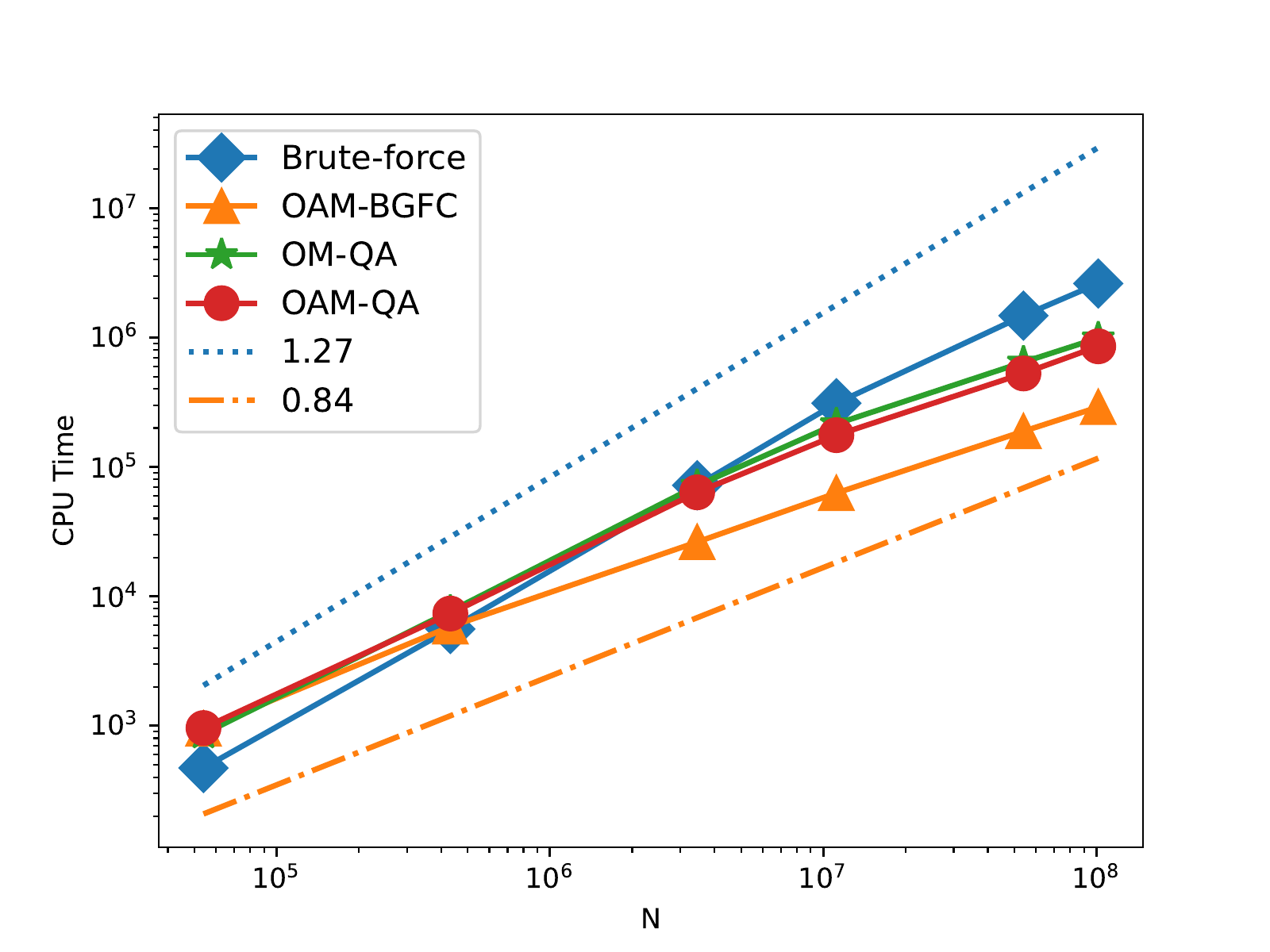}}
	\quad
	\subfigure[Time ratio]{
	\label{crack-time-ratio}
	\includegraphics[height=6cm]{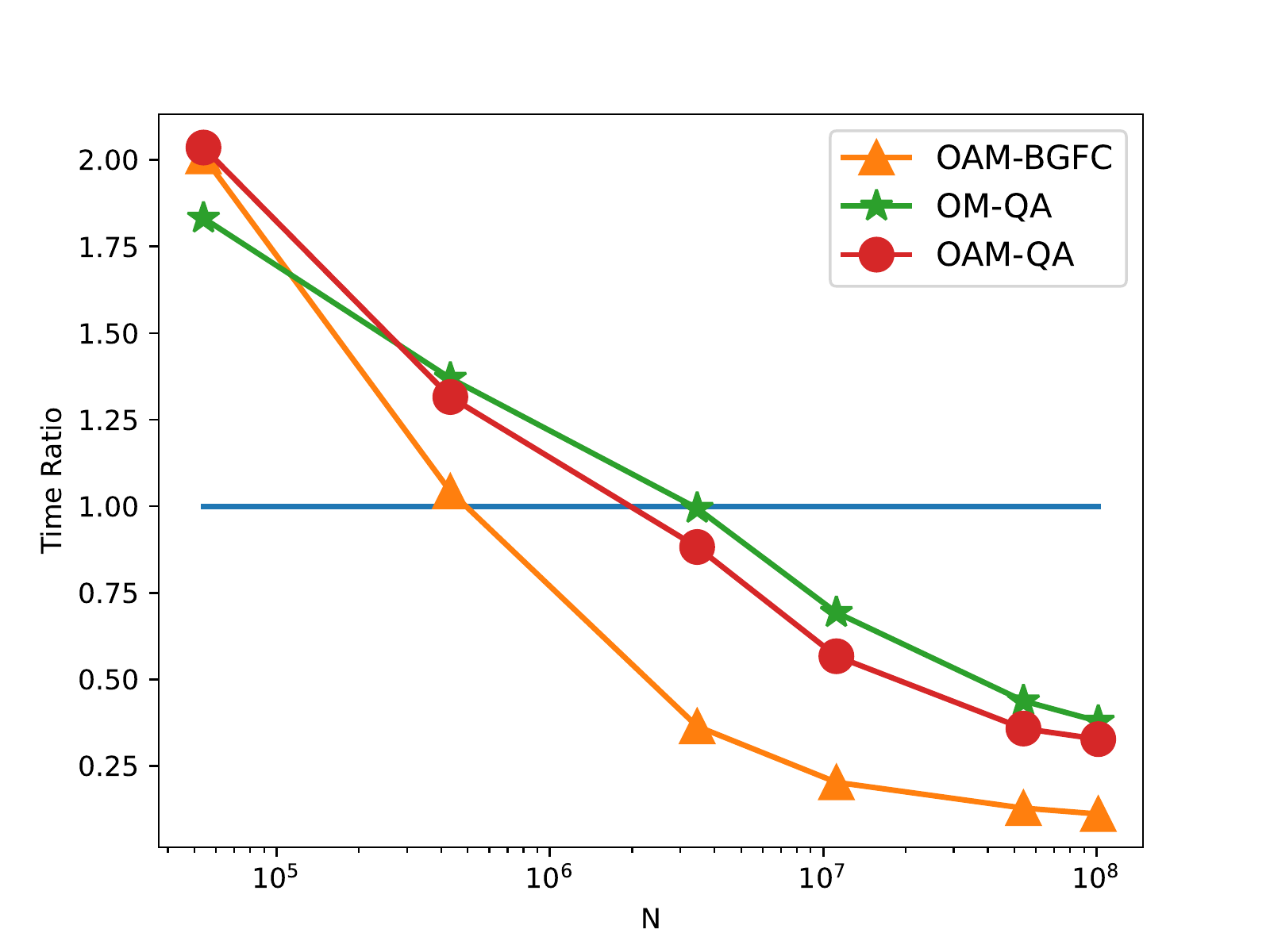}}
	\caption{CPU times of the OAM-BGFC approximation and its comparison with the brute-force optimization and oneway multigrid with other types of approximations, for the micro-crack case.}
	\label{fig:crack-time}
\end{figure}

Figure \ref{crack-cpu-time} presents the CPU times of the {\it brute-force} optimization (blue line), the oneway multigrid method (OM-QA, green line) proposed in \cite{chen2017efficient}, the oneway adaptive multigrid method with quasi-atomistic (OAM-QA, red line), and the oneway adaptive multigrid with BGFC method (OAM-BGFC, orange line) with respect to $N$ in the log–log scale. Figure \ref{crack-time-ratio} shows the ratio of CPU times between these methods and the {\it brute-force} optimization. 
Due to the increased non-locality of the displacement field in the presence of micro-crack, the brute-force optimization scales $O(N^{1.27})$ asymptotically. 

The green (OM-QA) and the red (OAM-QA) lines, representing the oneway multigrid with QA approximation, both scale a bit better $O(N^{1.10})$. The only difference between these two lines is the application of the local adaptive mesh (cf. Algorithm \ref{refine}). The green line is consistent with the numerical results in \cite{chen2017efficient}, while the red line indicates that the adaptivity only improves the pre-factor of the cost. 
Sublinear scaling ($O(N^{0.84})$) can still be observed if the OAM-BGFC method is considered. Moreover, compared with the {\it brute-force} optimization, the savings of the OAM-BGFC are around 80\% for systems with a hundred millions atoms. 
Note that the micro-crack considered in this paper is essentially a point defect, the adaptive mesh evolution of BGFC method for micro-crack is similar to the results for single vacancy (cf. Figure \ref{fig:sing-adpt-mesh-new-bgfc}). Hence, we omit it here for conciseness.

\subsection{Edge dislocation}
\label{sec:ed}

In the end, we consider the case of edge dislocation. We construct the (001)[100] edge dislocation in tungsten. It usually takes a quasi-2D manner \cite{mazars2011long}, that is, clamped boundary conditions are applied on the x-y plane while periodic boundary condition is imposed in the z-direction. In this paper, we aim to test our three dimensional implementation and apply the clamped boundary conditions in all three directions, with the solution of the quasi-2d problem as the boundary condition in the z-direction. 
Three layers of ghost atoms are utilized to fit the clamped boundary conditions. To obtain a better performance of the geometry optimization, the layers of ghost atoms on x-y plane have been relaxed. The reference atomistic systems contain from $3.305\times 10^5$ to $1.03\times 10^7$ atoms ({\it dofs}).

\begin{figure}[htb]
	\centering 
	\subfigure[Displacement $|u|$]{
	\label{dislocation-u}
	\includegraphics[height=6cm]{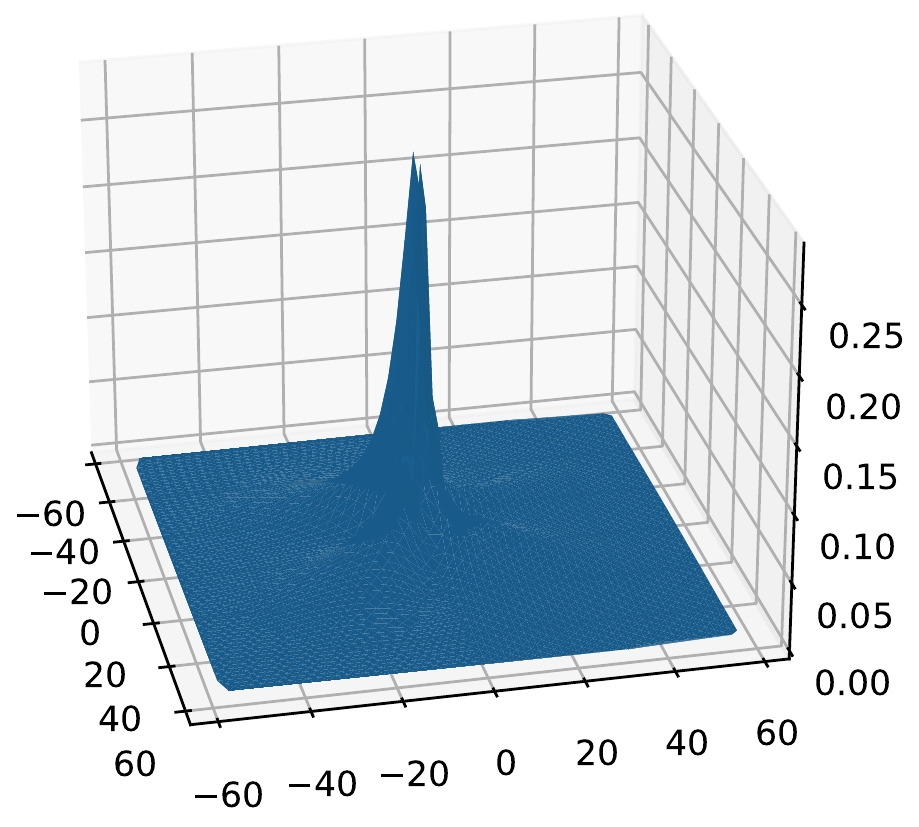}}
	\quad
	\subfigure[Decay of $|u(\ell)|$]{
	\label{dislocation-decay}
	\includegraphics[height=6cm]{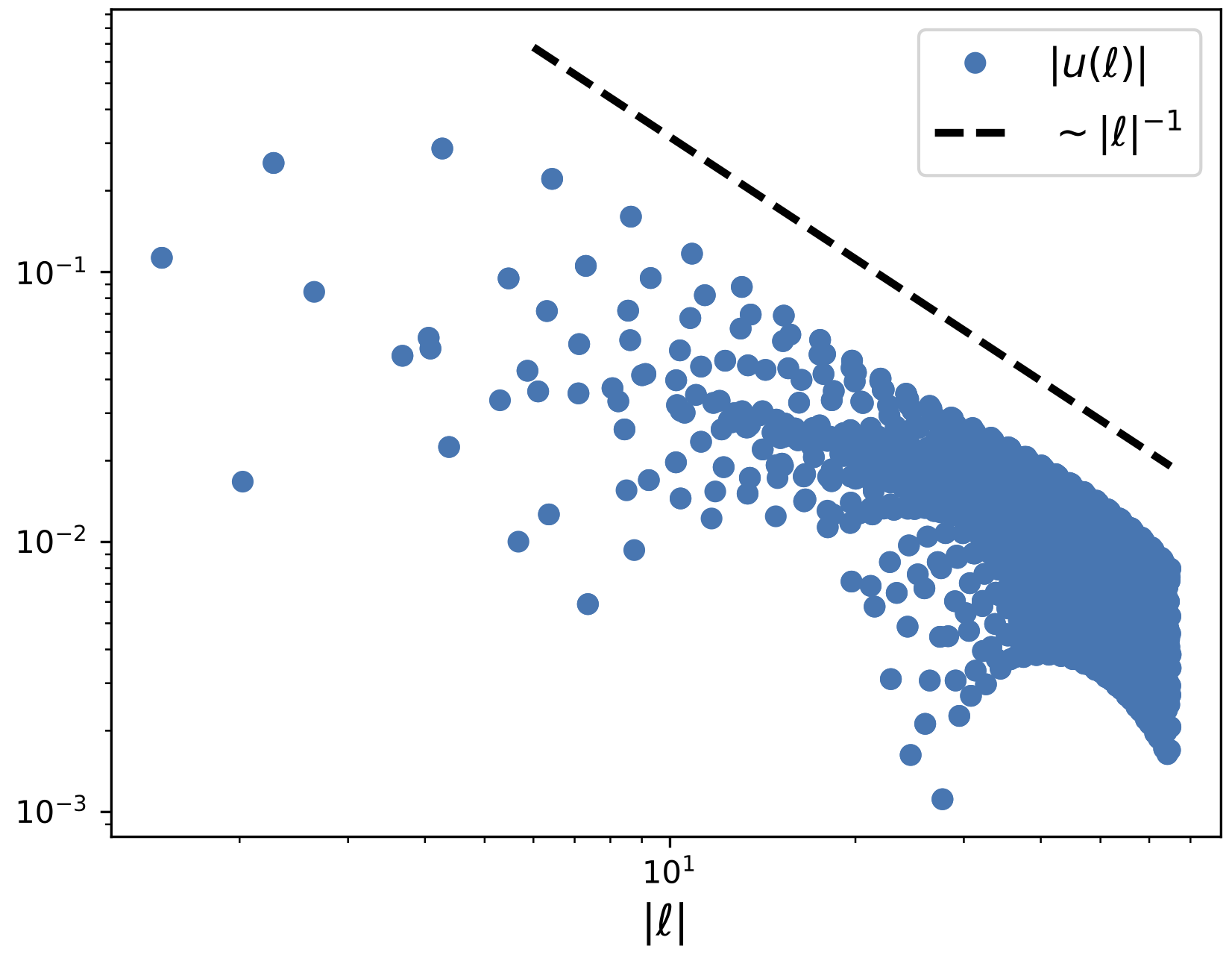}}
	\caption{The displacement and the decay of equilibrium for edge dislocation.}
	\label{fig:dislocation-u}
\end{figure}

Figure \ref{dislocation-u} plots the displacement field over a centered slice in the $z$ direction of the $30^3$ system. One can observe that in the presence of dislocations, the displacement field is less localized compared with point defects and typically decays like $|\ell|^{-1}$ \cite{anderson2017theory}, namely $|u(\ell)|\lesssim |\ell|^{-1}$, where $|\ell|$ is the distance to the defect core, which is also verified by Figure \ref{dislocation-decay}. Also, compared to the point defect cases (single vacancy and micro-crack), it is much more difficult to solve the geometry optimization problem for dislocations.

According to the numerical observations and discussions for single vacancy and micro-crack cases, the OAM-BGFC scheme is of particular practical interest and is much more efficient than that with the QA approximation. We therefore only test the OAM-BGFC method here.

\begin{figure}[H]
	\centering 
	\subfigure[CPU time]{
	\label{dislocation-cpu-time}
	\includegraphics[height=6cm]{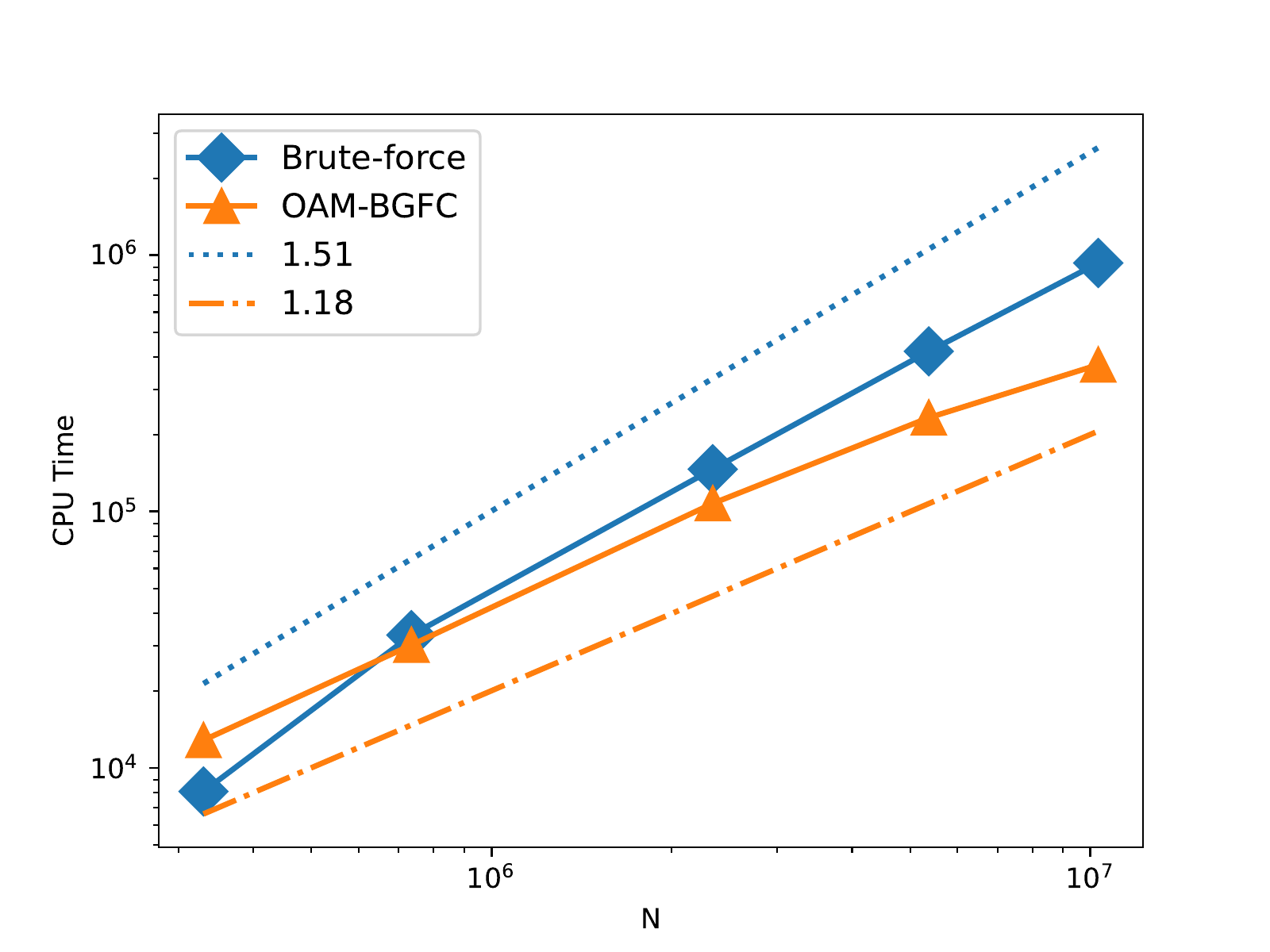}}
	\hskip 0.2cm
	\subfigure[Time ratio]{
	\label{dislocation-time-ratio}
	\includegraphics[height=6cm]{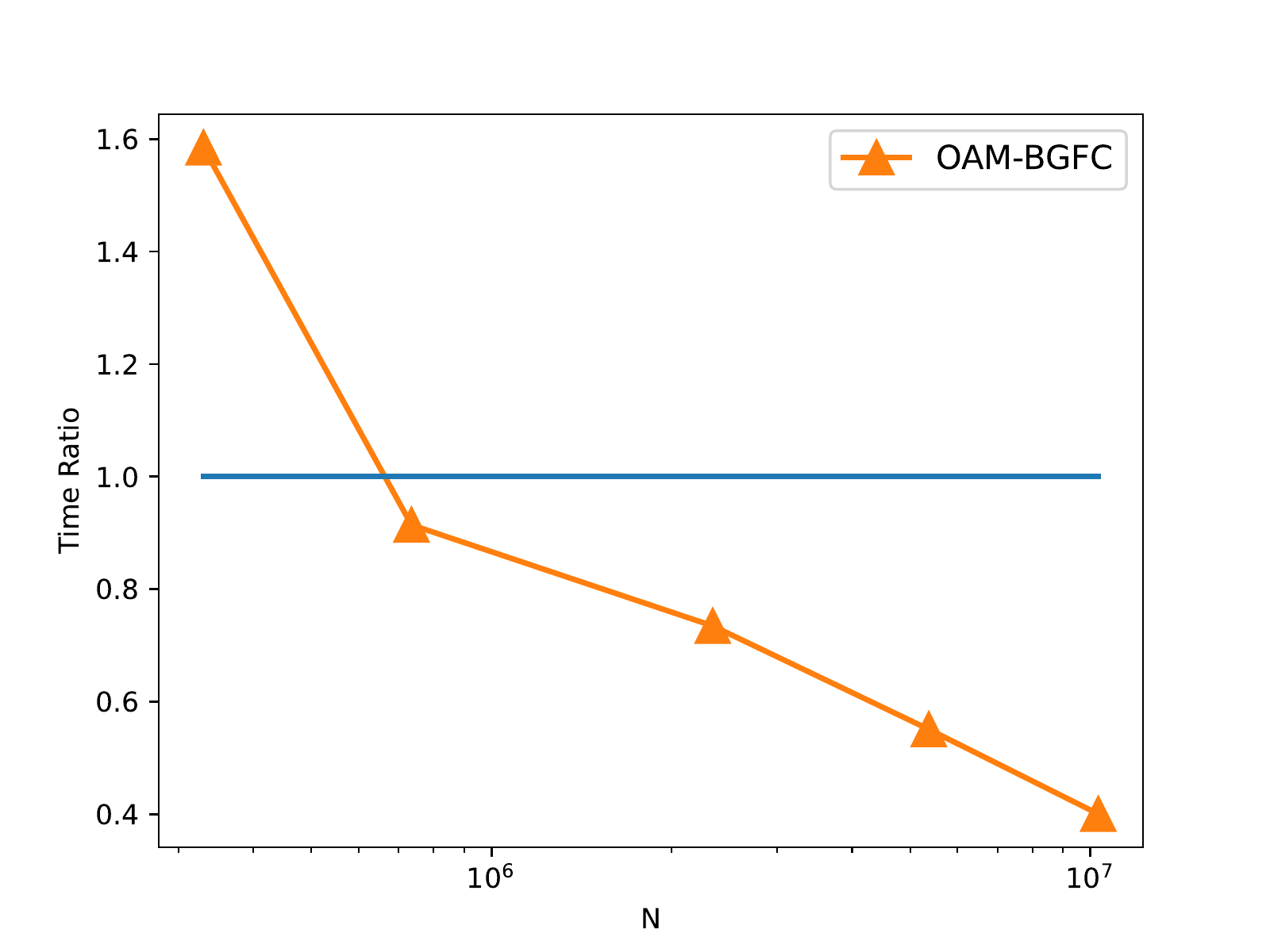}}
	\caption{CPU times of the oneway adaptive multigrid method with BGFC approximation (OAM-BGFC) and its comparison with the brute-force optimization for dislocation.}
	\label{fig:dislocation-time}
\end{figure}

Figure \ref{dislocation-cpu-time} presents the CPU times of the {\it brute-force} optimization (blue line) and the oneway adaptive multigrid with BGFC method (orange line) vs. $N$ in the log–log scale. Figure \ref{dislocation-time-ratio} shows the corresponding time ratio. 
As discussed above, due to the slower decay of the displacement field in the presence of edge-dislocation, the OAM-BGFC method scales near linearly ($O(N^{1.18})$) while the {\it brute-force} optimization scales $O(N^{1.51})$ asymptotically. Moreover, compared with the {\it brute-force} optimization, the savings of the OAM-BGFC method are around 60\% for systems with ten millions atoms. 

\begin{remark}\label{re:u0}
The superlinear scaling of the OAM-BGFC method for the edge dislocation case is probably due to the fact that we take the trivial $\hat{u}_0=0$ in \eqref{eq:gfc}. It is reasonable to speculate that a better choice of $\hat{u}_0$ may lead to an improved convergence rate for the BGFC approximation, and as a consequence, the corresponding OAM-BGFC method may achieve sublinear scaling. This will be investigated in depth in our future work.
\end{remark}

\begin{figure}[H]
	\centering 
	\includegraphics[height=6cm]{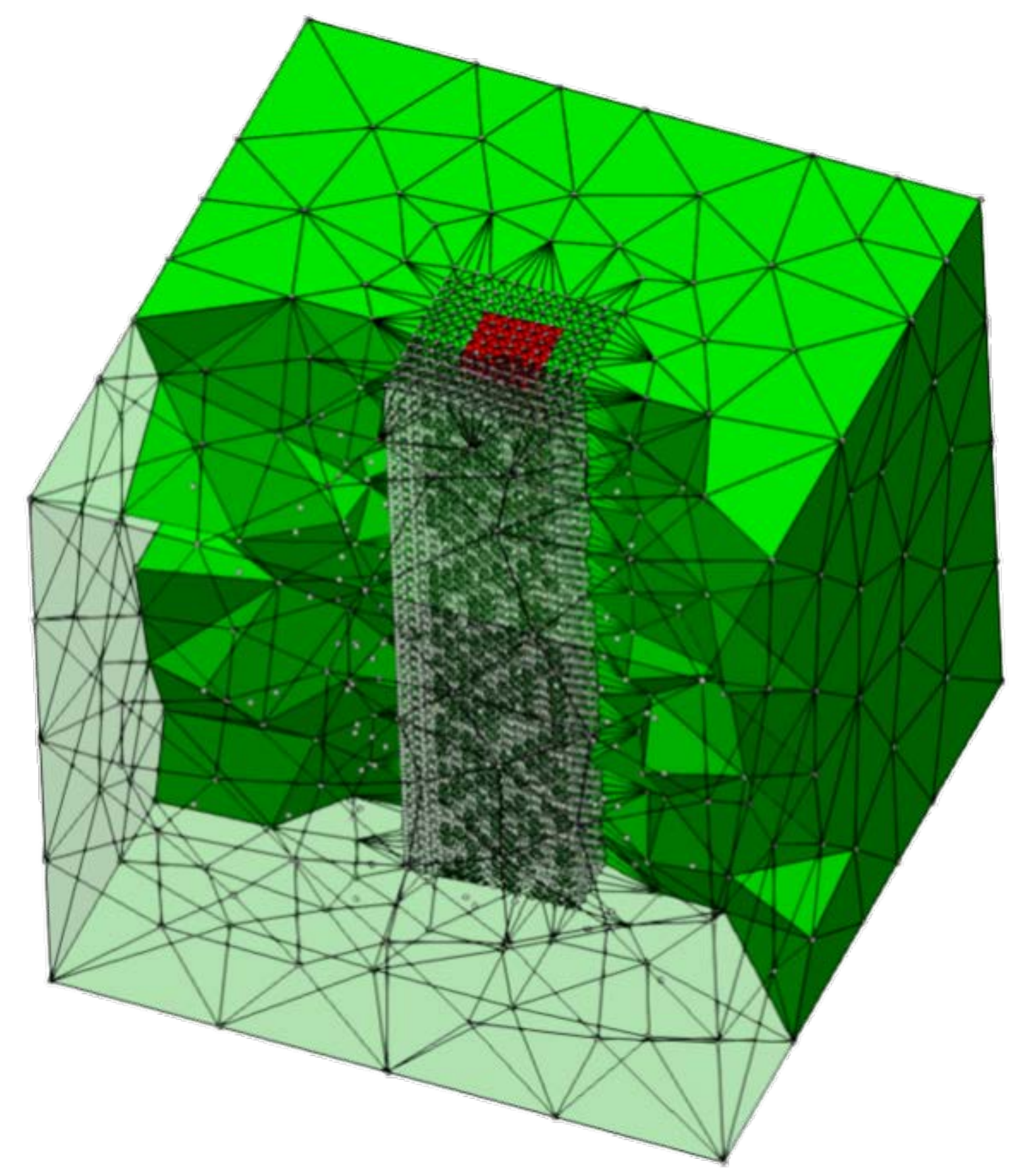}
	~~
	\includegraphics[height=6cm]{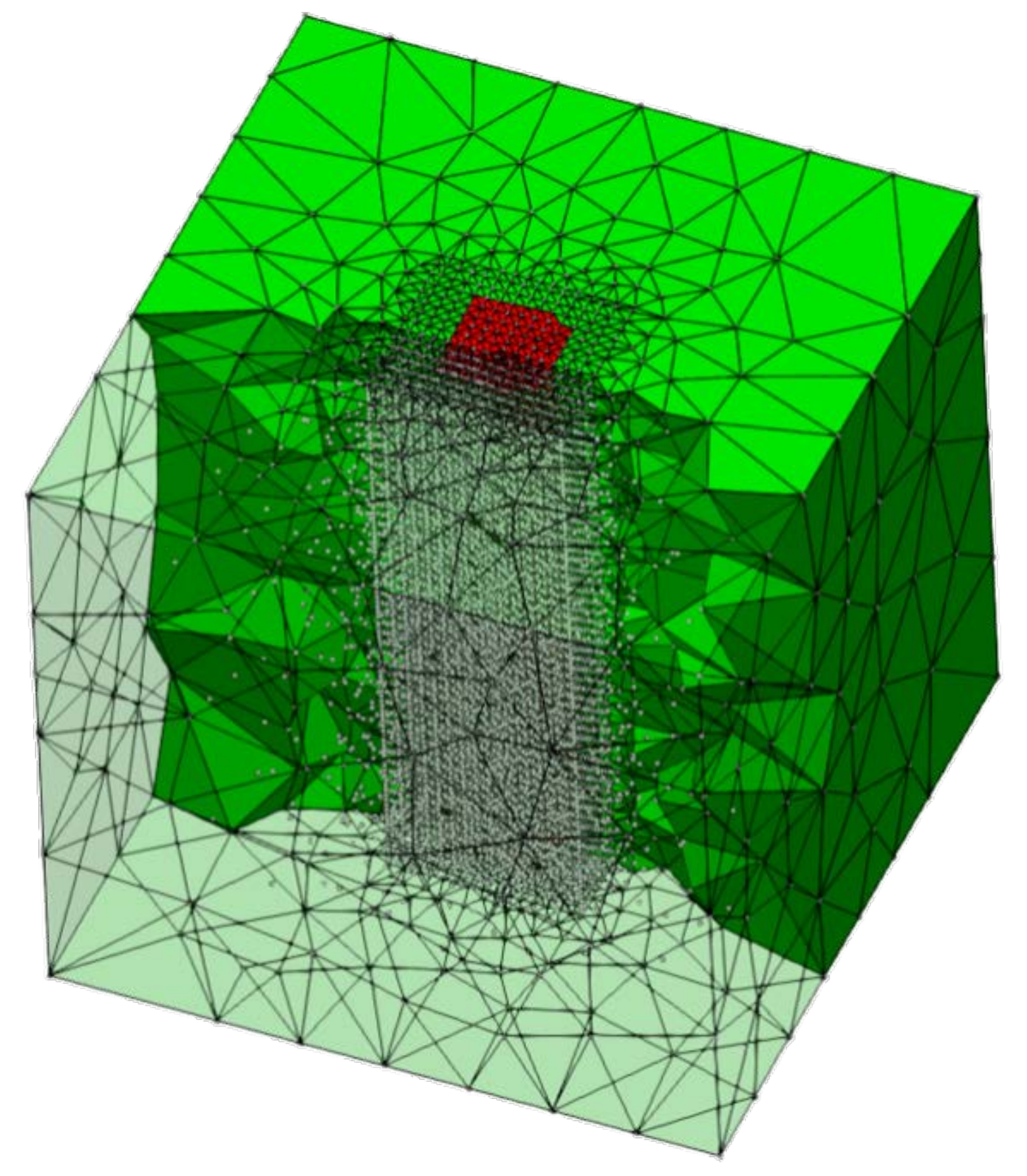}
	\\
	\includegraphics[height=6cm]{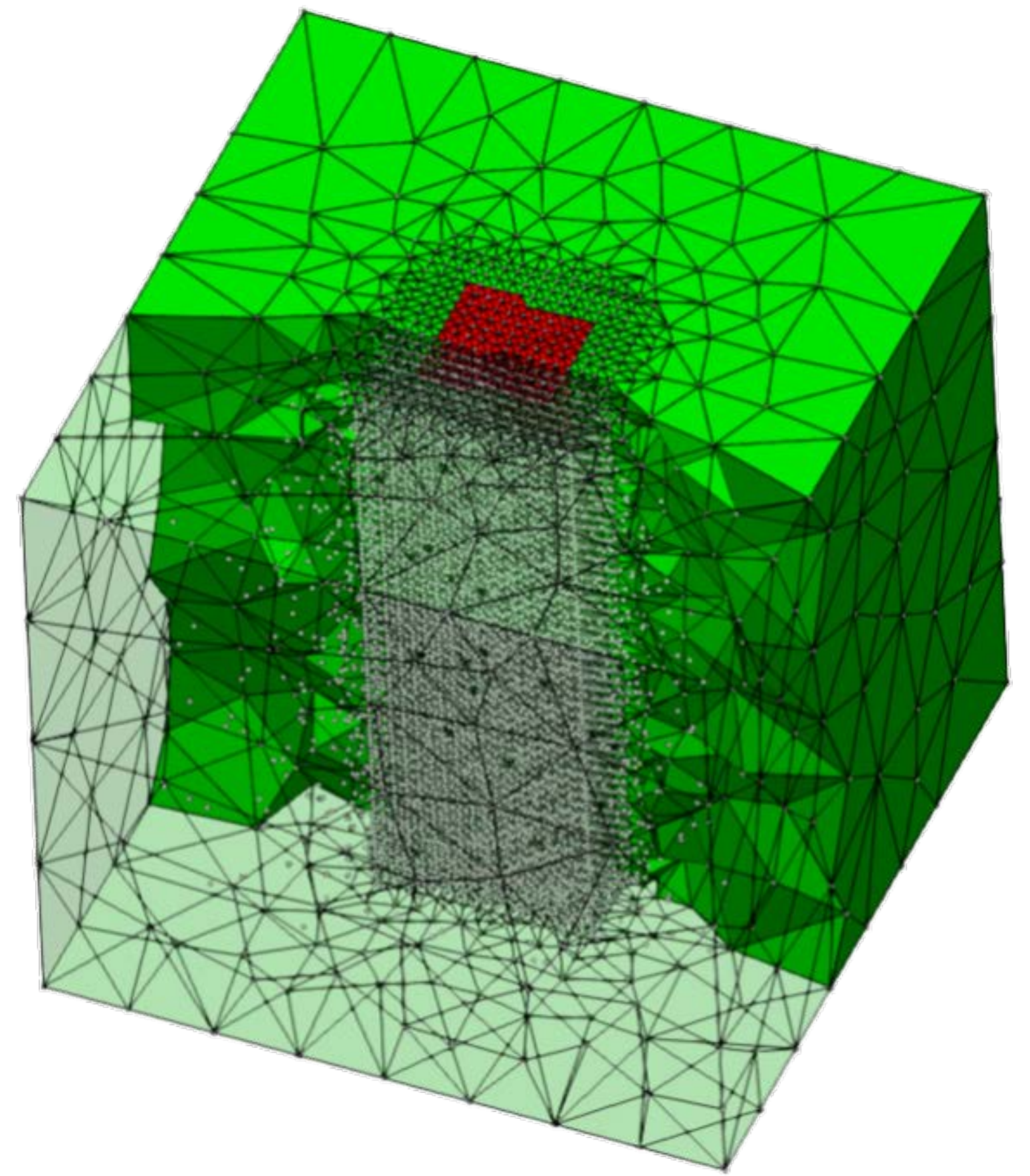}
	~~
	\includegraphics[height=6cm]{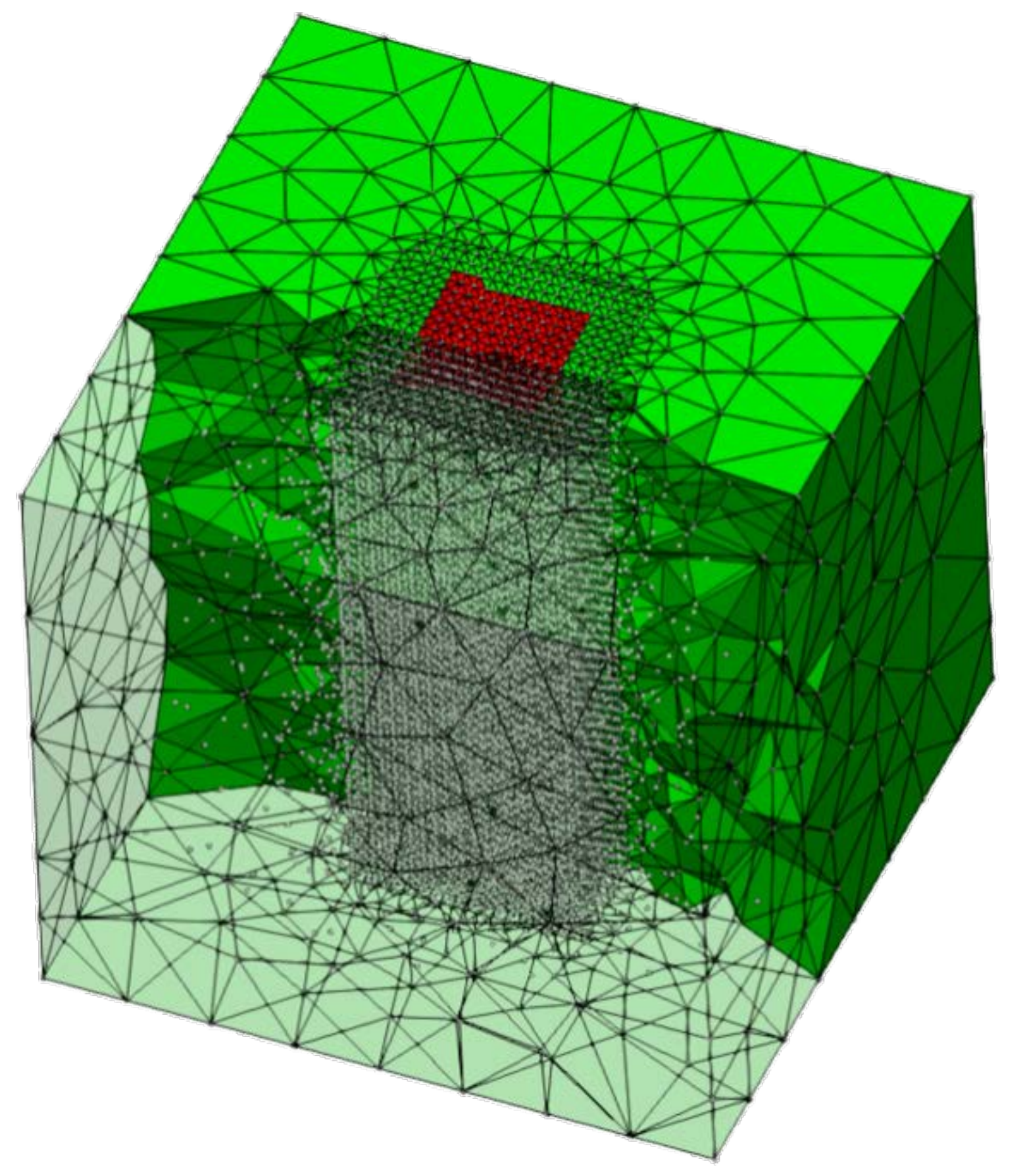}
 	\caption{Adaptive mesh refinement (degrees of freedom from upper left to lower right: 12667, 19913, 25863, 32377) of the BGFC approximation in OAM-BGFC, for edge dislocation. }
	\label{fig:dislocation-adpt-mesh}
\end{figure}

In Figure \ref{fig:dislocation-adpt-mesh} we plot the mesh evolution in the adaptive process for the OAM-BGFC method for the $30^3$ system. 

\section{Conclusions}
\label{sec:conclusion}

We develop an efficient adaptive multigrid strategy for the geometry optimization of large-scale three dimensional molecular mechanics. The resulting method can achieve an optimal near linear or even sublinear computational scaling by exploiting the intrinsic low-rank structure for local defect configurations such as vacancies and dislocations. We propose a oneway multigrid method with adaptive blended ghost force correction (OAM-BGFC) approximation on the coarse levels, combined with adaptive mesh refinements based on some gradient-based {\it a posteriori} error estimators. We utilize state-of-the-art 3D mesh generation techniques to effectively implement the algorithm. For systems with up to a hundred million atoms, this strategy has a five-fold acceleration in terms of CPU time compared with the {\it brutal force} optimization. 

Although we believe such strategy is generally applicable to other multiscale coupling methods and more complex crystalline defects, this research still raise a few open problems which deserve further mathematical analysis and algorithmic developments. 

\begin{enumerate}
    \item 
The first one is the adoption of more complex multigrid strategies like full multigrid (FMG) and full approximation scheme (FAS) in the geometry optimization of molecular mechanics. We believe that the ideas in this paper makes it possible to develop more efficient algorithms. 
    \item 
The second one is the problem of the rigorous {\it a priori} error estimate of BGFC approximation for straight dislocations. A possible remedy is to study an equivalent ghost force removal formulation, where the construction of a suitable ``predictor" $\hat{u}_0$ is essential, as discussed in Remark \ref{re:disloc} and Remark \ref{re:u0}. Once the {\it a priori} error estimate is developed, the study of the {\it a posteriori} error estimation and the corresponding adaptive algorithm should be underway, where \cite{wang2018posteriori, wang2021posteriori} should provide good references. 
    \item 
The last but may be the most important problem to consider is the adaptive multigrid strategies for complex crystalline defects including cracks, partial dislocations and grain-boundaries, which have already attracted considerable attentions. The main bottleneck is the construction and implementation of the corresponding a/c coupling approximations.
\end{enumerate}

\section*{Acknowledgments}
ML and LZ were partially supported by National Natural Science Foundation of China (NSFC 11871339, 11861131004). YW is supported by Development Postdoctoral Scholarship for Outstanding Doctoral Graduates from Shanghai Jiao Tong University. 

\appendix
\renewcommand\thesection{\appendixname~\Alph{section}}

\section{Quasi-atomistic (QA) approximation}
\label{sec:appendix}
\renewcommand{\theequation}{A.\arabic{equation}}
\renewcommand{\thefigure}{A.\arabic{figure}}
\renewcommand{\thealgorithm}{A.\arabic{algorithm}}
\setcounter{equation}{0}
\setcounter{figure}{0}
\setcounter{algorithm}{0}

We briefly introduce the mesh generation and the oneway adaptive multigrid algorithm for the QA approximation (cf. \S~\ref{sec:qa} and \cite{chen2017efficient}).

\begin{figure}[htb]
	\centering 
	\includegraphics[height=6cm]{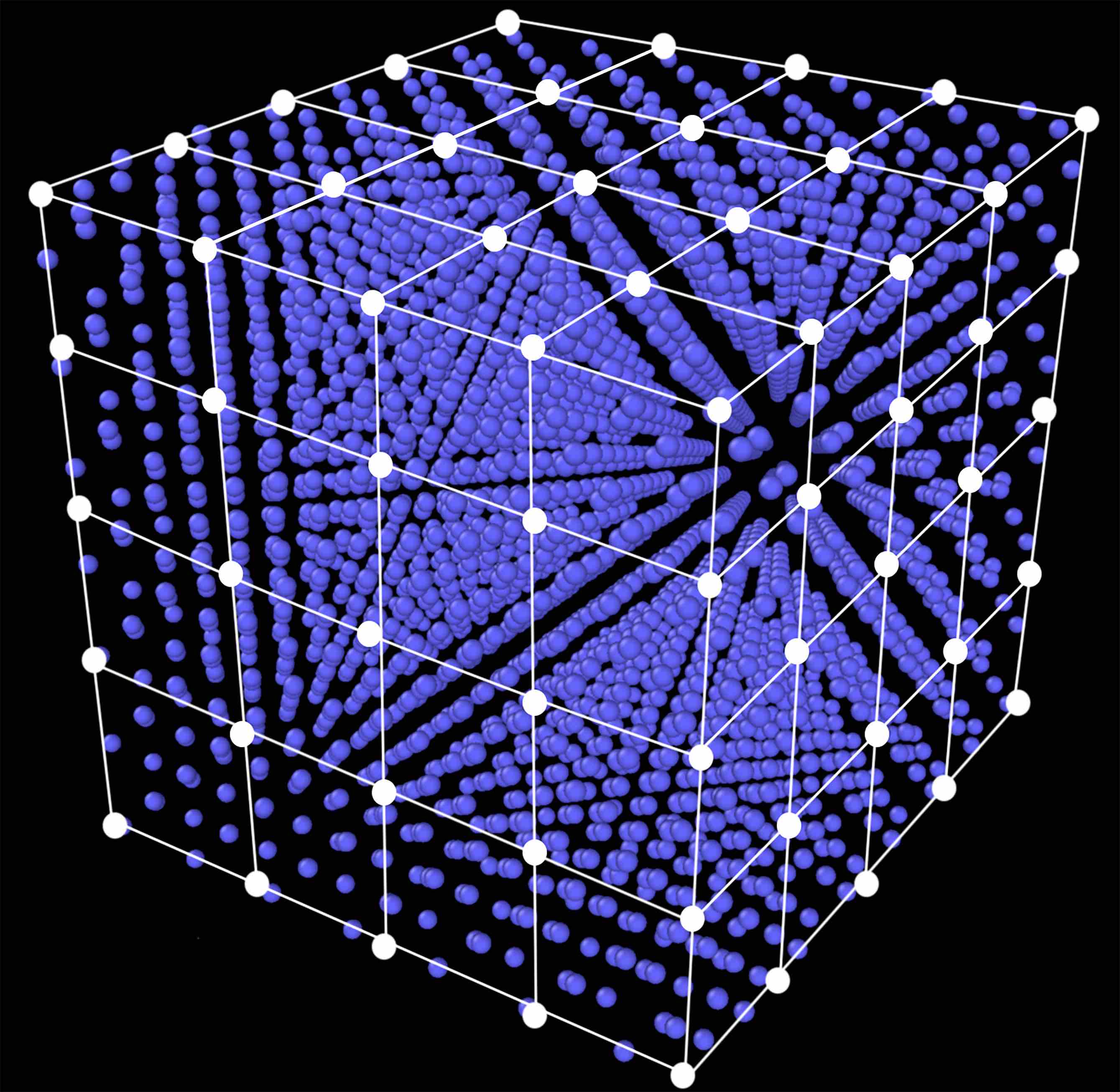}
	\caption{Illustration of the cubic triangulation generated by {\tt deal.II} for quasi-atomistic approximation.}
	\label{fig:mesh}
\end{figure}

\begin{figure}[H]
	\centering 
	\includegraphics[scale=0.16]{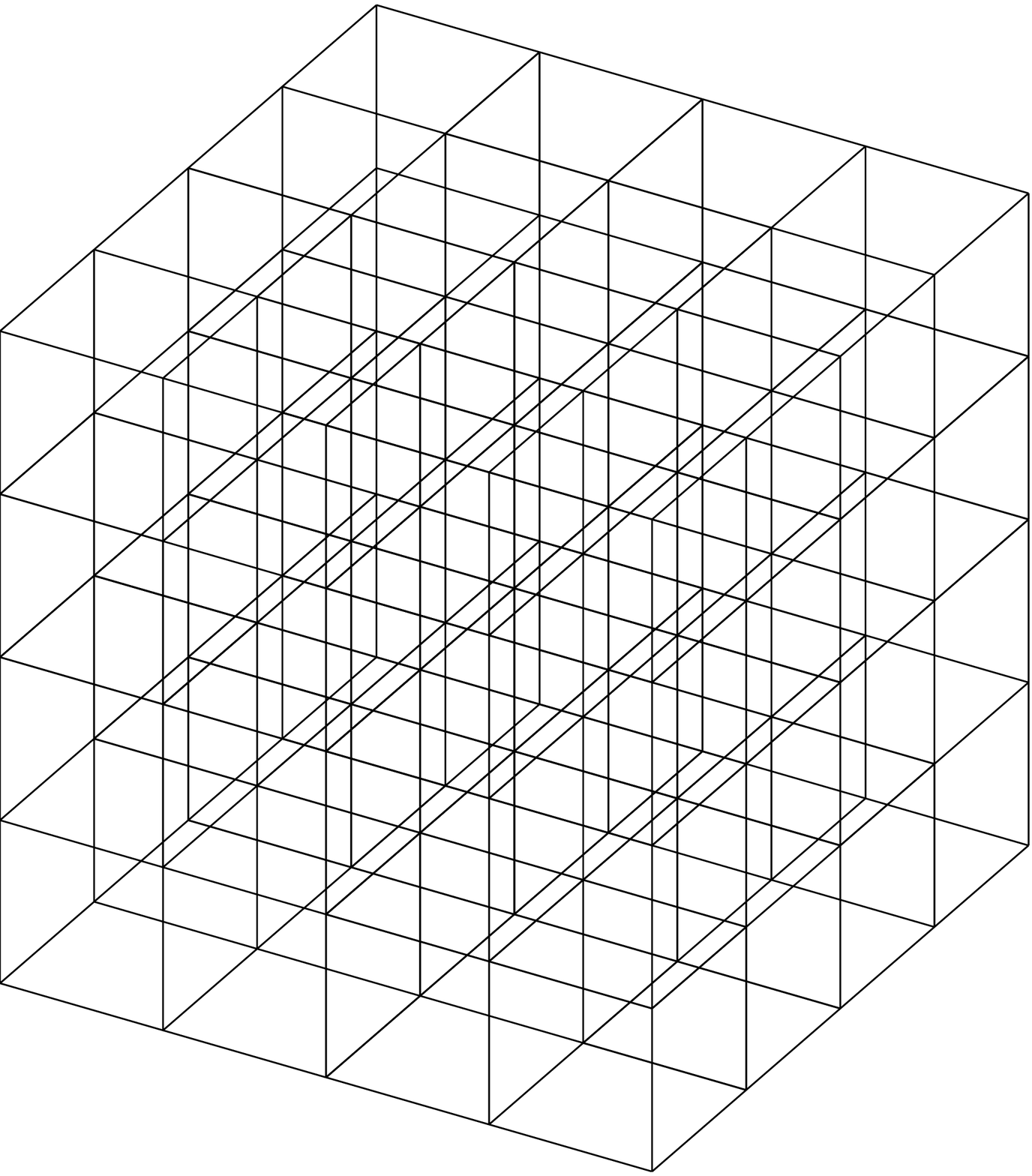}
	~~
	\includegraphics[scale=0.16]{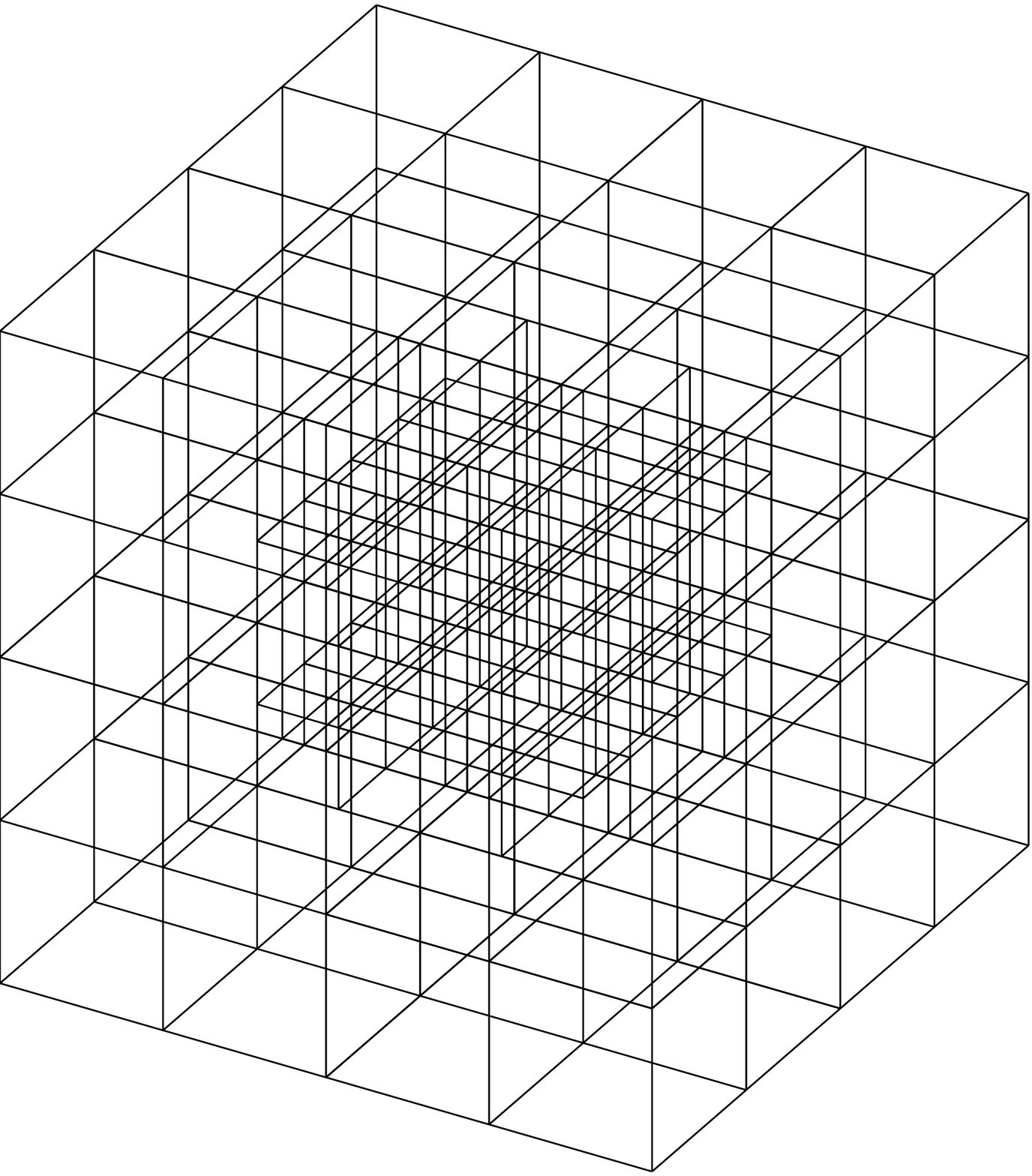}
	~~
	\includegraphics[scale=0.16]{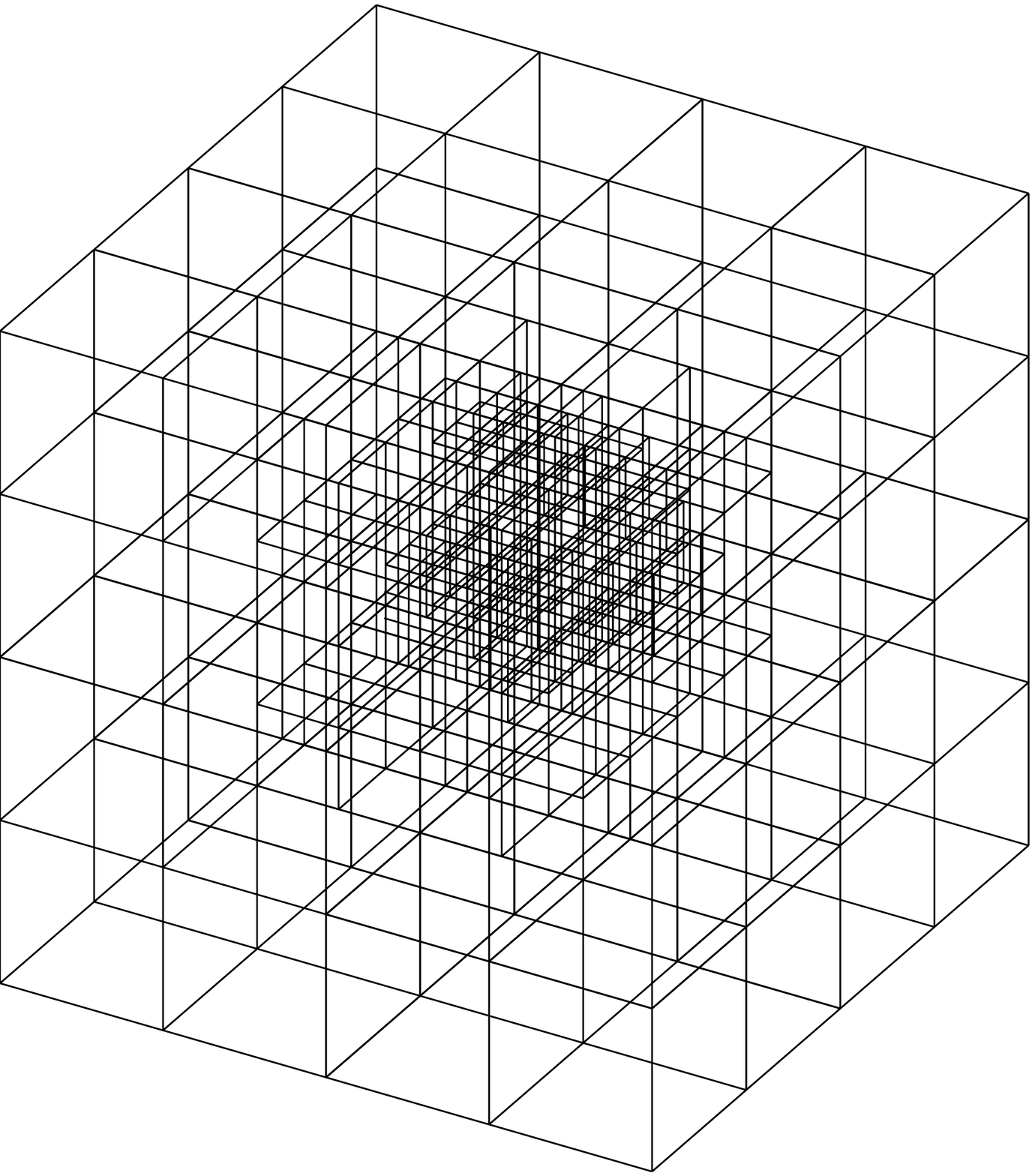}
	~~
	\includegraphics[scale=0.16]{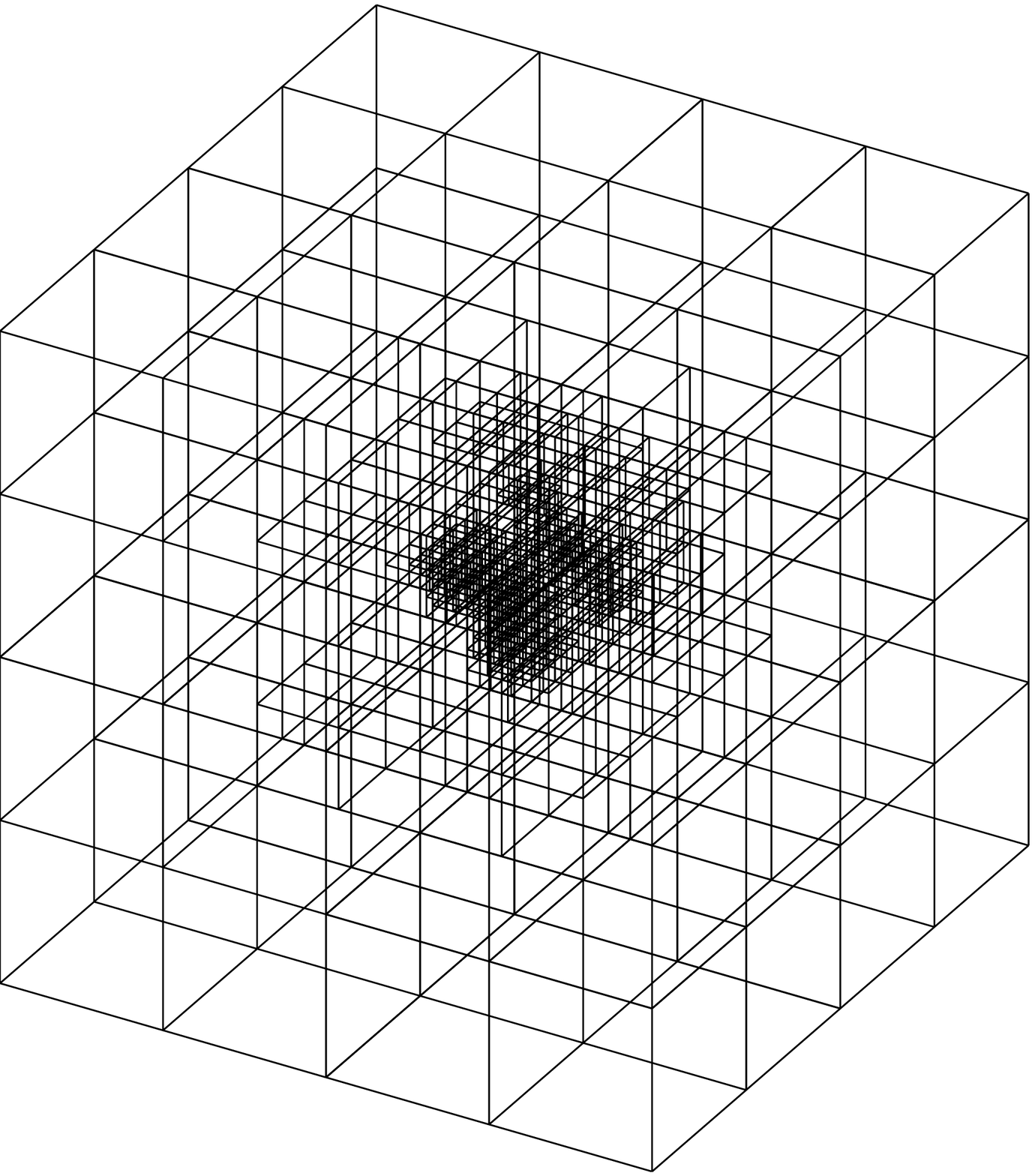}
 	\caption{Adaptive mesh refinement ({\it dofs} from left to right: 125, 272, 618, 1536) of the QA approximation for a single vacancy.}
	\label{fig:sing-adpt-mesh-new}
\end{figure}

As we mentioned in \S~\ref{sec:at}, the uniform cubic triangulation was used in \cite{chen2017efficient} when the QA approximation is employed as the coarse-grid problem. See Figure \ref{fig:mesh} for an illustration. To make a direct comparison, we still apply the same cubic triangulation but adapt it to combine with the local adaptive mesh refinement (cf. Figure~\ref{fig:sing-adpt-mesh-new}).

We use the package \texttt{deal.II} \cite{dealII90, BangerthHartmannKanschat2007} to generate the cubic triangulation. \texttt{deal.II} is open-source and well-developed, and can deal with the {\it hanging nodes} automatically \cite{dealII90, BangerthHartmannKanschat2007}, which is a well-known difficulty in the adaptation of the cubic triangulation.
The local adaptive mesh refinement can also be achieved by \texttt{deal.II} once the local error estimators are given, which has already been discussed in \S~\ref{sec:algo}.

Next, similar to Algorithm \ref{alg:refine}, we introduce the mesh refinement strategy for the QA approximation as follows. Compared with Algorithm \ref{alg:refine}, where a suitable strategy to  enlarge the atomistic region is included, Algorithm \ref{refine} contains only the mesh refinement, which is implemented by using the utilities in \texttt{deal.II} directly. 

\begin{algorithm}[H]
\caption{Mesh refinement strategy for QA approximation.}
\label{refine}
\quad Prescribe $0<\tau_1<\tau_2<1$ and $k \in  \N$.
\begin{enumerate}
	\item  Given a partition $\T_k$, according to the approximate solution $u_k$, compute the local error estimator $\eta_T = ||\nabla u_k||_{L^{2}(T)}$ for each element $T \in \T_k$ and the global error estimator $\eta_{\T_k} = \sum_{T \in \T_k} \eta_T$.
	\item  Choose the maximum sets $\T_{\rm c} \subset \T_k$ and the minimum sets $\T_{\rm r} \subset \T_k$ such that the following D\"{o}rfler properties are satisfied
	\begin{equation}
	\sum_{T \in \T_{\rm c}} \eta_T \leq \tau_1 \eta_{\T_k}, \quad \sum_{T \in \T_{\rm r}} \eta_T \geq \tau_2 \eta_{\T_k}.
	\end{equation}
	\item  Mark all the elements in $\T_{\rm r}$ for refinement (bisection) and the elements in $\T_{\rm c}$ for coarsening to obtain $\T_{k+1}$.
\end{enumerate}
\end{algorithm}

We then present the following adaptive multigrid strategy.

\begin{algorithm}[H]
\caption{Oneway adaptive multigrid strategy with QA approximation (OAM-QA).}
\label{alg:qa}
\begin{enumerate}
	\item Prescribe the parameters $\mu_k$ and $c$.
	\item Relax the QA problem $\mathbf{P}^{\rm QA}_0$ defined by \eqref{eq::uka} on the initial mesh $\T_0$ for $\mu_0$ times to obtain $u_0$ with a trivial initial guess, set $k=1$.
	\item Carry out the mesh refinement based on Algorithm \ref{refine} to obtain $\T_k$, relax the quasi-atomistic problem $\mathbf{P}^{\rm QA}_k
	$ defined by \eqref{eq::uka} on $\T_k$ for $\mu_k$ times to obtain $u_k^{{\rm qa}, (\mu_k)}$ with the initial guess $I^k_{k-1}u^{{\rm qa}, (\mu_{k-1})}_{k-1}$, compute $h_k$. If $h_k \leq ch_{N}$, $k=k+1$, repeat Step 3; Otherwise, goto Step 4. 
	\item Solve the atomistic problem $\mathbf{P}^{\rm A}_N$ defined by \eqref{eq::a} until it is convergence with the initial guess interpolated from the solution on $\T_{k}$.
	\end{enumerate}
\end{algorithm}

We briefly discuss the stopping criterion in Algorithm \ref{alg:qa}. As \cite{chen2017efficient} shows, any attempt to solve the coarse-grid model (QA approximation) to high accuracy will slow down the convergence of multigrid overall. Hence, only a few iterations are required at each level. Throughout the numerical experiments presented in this paper, we set $\mu_k=3$ in Algorithm \ref{alg:qa}, which is the same setting as that in \cite{chen2017efficient}.

The mesh evolution in the adaptive process for the oneway adaptive multigrid with QA approximation for single vacancy introduced in \S~\ref{sec:sv} is given as follows. 



\bibliographystyle{elsarticle-num} 
\bibliography{AdapMG.bib}

\end{document}

%% file: notation.tex
\def\XXint#1#2#3{{\setbox0=\hbox{$#1{#2#3}{\int}$ }
\vcenter{\hbox{$#2#3$ }}\kern-.6\wd0}}

\def\b{\big}

\DeclareMathOperator*{\argmin}{argmin}

\def\R{\mathbb{R}}
\def\N{\mathbb{N}}
\def\Z{\mathbb{Z}}

\def\dx{\,{\textrm{d}}x}

\def\<{\langle}
\def\>{\rangle}

\def\mA{{\textsf{A}}}

\def\mF{\textsf{F}}

\def\mI{\textsf{I}}

\def\a{\textrm{a}}
\def\c{\textrm{c}}
\def\b{\textrm{b}}

\def\L{\Lambda}

\def\Us{\mathscr{U}}

\def\E{\mathscr{E}}

\def\L{\Lambda}

\def\Nhd{\mathcal{N}}
\def\rcut{r_{\textrm{cut}}}
\def\Rg{\mathscr{R}}

\def\T{\mathcal{T}}

\def\T{\mathcal{T}}

\def\vor\textrm{vor}

\def\Rg{\mathcal{R}}

\def\rcut{r_{\textrm{cut}}}
\def\Lhom{\L^{\textrm{hom}}}

\def\Rg{\mathcal{R}}

\def\rcut{r_{\textrm{cut}}}
\def\Lhom{\L^{\textrm{hom}}}

\def\vor{\textrm{vor}}
